\documentclass[twocolumn,showpacs,superscriptaddress,prb]{revtex4-1}
\usepackage{graphicx}
\usepackage{color}
\usepackage{amsmath,amsthm,amssymb}
\usepackage{braket}
\usepackage[colorlinks,citecolor=blue,linkcolor=red]{hyperref}
\usepackage{multirow}
\begin{document}
 
\title{Identifying Topological Superconductivity in 2D Transition-Metal Dichalcogenides}

\author{Christopher Lane}
\email{laneca@lanl.gov}
\affiliation{Theoretical Division, Los Alamos National Laboratory, Los Alamos, New Mexico 87545, USA}
\affiliation{Center for Integrated Nanotechnologies, Los Alamos National Laboratory, Los Alamos, New Mexico 87545, USA}

\author{Jian-Xin Zhu}
\email{jxzhu@lanl.gov}
\affiliation{Theoretical Division, Los Alamos National Laboratory, Los Alamos, New Mexico 87545, USA}
\affiliation{Center for Integrated Nanotechnologies, Los Alamos National Laboratory, Los Alamos, New Mexico 87545, USA}

\date{\today} 
\begin{abstract}
We study the superconducting pairing instabilities and gap functions for prototypical two-dimensional (2D) transition-metal dichalcogenides (TMDCs) WS$_2$, MoTe$_2$, and MoS$_2$ in the 2H phase under both hole and electron doping at $10$ K. Our first-principles quantum many-body Green’s function approach allows us to treat the full $d$ and $p$ manifold of orbitals with strong spin-orbit coupling, yielding pairing predictions with material specific detail. The resulting gap functions exhibit a variety of mixed-parity superconducting states, including $s$, $p$, $d$, $f$, $d\pm id$, and $p\pm ip$ pairing modes. In particular, we predict 3\% and 4\% hole-doped WS$_2$ to be a chiral $p\pm ip$ topological superconductor. For 1\% hole-doped MoS$_2$, we find a competition between three doubly degenerate  chiral and non-chiral instabilities. Overall, the relative pairing strengths are found to follow the Fermi surface topology, due to nesting between the Fermi surface sheets. Finally, we discuss our predictions in relation to available experimental data and classify the topology of the predicted superconducting pairing symmetries.
\end{abstract}


\maketitle 

\section{Introduction}
Spurred by the promise of quantum information processing that outperforms classical supercomputers, significant advances have been made in realizing quantum supremacy. In just the last few years, quantum computational advantage\cite{preskill2012quantum},  has been successively achieved using superconducting Josephson junctions (Sycamore)\cite{arute2019quantum} and photonic-based (Jiuzhang and Zuchongzhi) \cite{zhong2020quantum,wu2021strong} quantum processors. Other demonstrations include implementing quantum approximate optimization algorithms on trapped-ion processors\cite{pagano2020quantum} and using bosonic modes to perform computationally hard problems\cite{huh2015boson}. 

With our entrance into the noisy intermediate-scale quantum (NISQ) era, the next frontier is to achieve quantum advantage for practical problems. For quantum based processors to deliver breakthroughs in currently intractable problems in genetics, chemistry and molecular dynamics, materials science, and data encryption, errors must be pushed to better than $1$ in $10^9$. However, despite recent advances, leading qubit architectures still suffer from  intrinsic limitations of coherence times $\sim1$ ms\cite{resch2019quantum,martonosi2019next} and an operating  fidelity of $\sim99.9$\%\cite{resch2019quantum,martonosi2019next, egan2020fault}, placing greater focus on quantum error mitigation to enable sustainable quantum supremacy. Multiple schemes have been introduced to spread the information of one qubit out over a few auxiliary qubits, thus making the calculation less susceptible to a single point fault. The simplest of these approaches employs seven\cite{shor1995scheme} to nine\cite{steane1996error} auxiliary qubits to encode one logical qubit. However, when scaled to hundreds or thousands of logical quibits, which is necessary for tackling hard practical problems, the overhead can easily become unmanageable even with the most modest errors rates. Therefore, other routes must be examined.

One such path forward is to use {\it topological} qubits. Topological qubits provide a path towards error-tolerant quantum computing by embedding quantum information into the global ground state properties of a system\cite{kitaev2003fault,nayak2008non}, thereby, making them intrinsically immune to decoherence stemming from spurious perturbations. Specifically, fault-tolerant quantum computations would be carried out by `braiding' non-Abelian anyons,  quasiparticles that obey non-Abelian statistics and accordingly are neither fermions nor bosons. As a consequence, these quantum processors inherently push errors to 1 in $10^7$\cite{gibney2016inside}, providing a much more reasonable starting point for applying any error mitigation methodologies. 

Currently, quantum spin liquids\cite{zhou2017quantum,clark2021quantum} and topological superconductors\cite{flensberg2021engineered,qi2011topological,sato2017topological,de2021materials} are the most intensely studied families of material systems for realizing topological qubits. In particular, Majorana bound states are the most promising non-Abelian topological quantum states for quantum computation and naturally arise in quantum spin liquids the fundamental fractional spin excitations, and in  topological superconductors, due to the native particle-hole symmetry and nontrivial topology. So far, very few material realizations have been theoretically predicted, let alone experimentally verified. 

Current experimental efforts in identifying topological superconductors have been centered on four families of materials: proximity induced and heterostructures\cite{lutchyn2018majorana}, complex oxides\cite{maeno2011evaluation}, doping and intercalation of topological insulators\cite{sasaki2011topological}, and heavy fermion compounds\cite{schemm2014observation}. 	Nevertheless, each material candidate possesses its own challenges such as required fabrication precision, dopant aggregation and inhomogeneity, or toxicity. This then begs the following question: are there `simple' material platforms with intrinsic topological superconductivity?

Atomically thin two-dimensional (2D) materials have proven to be one of the most exciting platforms exhibiting an extensive range of novel electronic\cite{kim2015observation}, excitonic\cite{mueller2018exciton}, valley\cite{hung2019direct}, topological\cite{choe2016understanding}, and charge density waves\cite{gye2019topological}. The 2D transition-metal dichalcogenides (TMDCs), in particular, encompass an expansive phase space of pristine compounds allowing for a range of spin-orbit-coupling strengths, $d$-electron counts,  and crystal structures. Furthermore, superconductivity has already been observed in a number of TMDCs either intrinsic or induced by chemical doping, electrostatic doping or applied pressure\cite{manzeli20172d,qiu2021recent}. But, the pairing symmetry and a complete microscopic mechanism still remains to be determined. Due to the strong spin-orbit coupling and the presence of superconductivity in many of the TMDCs, these materials provide a promising alternative route to engineer Majorana fermions in solid-state systems. 

Despite the advanced experimental efforts, robust theoretical predictions for new topological superconducting material platforms are still lacking. Presently, theoretical insights are typically given by limited low-energy models designed to extract specific physical properties for a general class of materials. If we wish to tackle even the `simplest' of materials, such as the TMDCs, the theoretical framework employed must be able to address the intertwining of strong spin-orbit coupling, multiple active local orbitals at the Fermi level, correlation effects, and lattice vibrations on the same footing. This is essential for a transparent theoretical examination of the microscopic mechanism of topological superconductivity in material specific detail.

In this paper we present a first-principles quantum many-body Green’s function approach to examine the mechanism of topological superconductivity in material specific detail. Starting from the two-particle Green's function, we systematically derive a spin-orbital dependent self-consistent expression for the effective pairing interaction and superconducting gap. This effective pairing potential follows in the spirit of the Kohn-Luttinger mechanism and is composed of a generalized set of random phase approximation (RPA) type equations. To analyze the resulting spin and orbital gap functions, we introduce a symmetry agnostic scheme so as to not introduce any biasing assumptions. We then apply this  treatment to WS$_2$ in detail since it displays a rich variety of Fermi surface topologies, strong spin-orbit coupling, and $d$-orbital character. Since the carriers in the TMDCs as a whole have strong $d$-orbital character, correlation effects are expected to be important. This has been confirmed by a number of experimental studies reporting magnetism in MoS$_2$ and various other TMDCs\cite{zhang2007magnetic,li2008mos2,mathew2012magnetism,ma2012evidence}, along with a recent study observing an unconventional scaling of the superfluid density with critical temperature similar to the high-Tc cuprates\cite{von2019unconventional}. Therefore, this study focuses on correlation driven superconductivity, with phonon effects to be added in future works since they are predicted to be weaker\cite{roldan2013interactions}. Finally we compare the results of three prototypical 2D TMDCs: WS$_2$, MoTe$_2$, and MoS$_2$, and discuss our prediction in relation to available experimental data and classify the topology of the predicted superconducting pairing symmetries.

The outline of this paper is as follows: In Sec.~\ref{sec:theory} the formal theoretical approach is laid out along with a summary of the computational details. In Sec.~\ref{sec:pairing} a detailed examination of the pairing instabilities and symmetries of WS$_2$ is presented for both electron- and hole-doping cases. In  Sec.~\ref{sec:compare_pairing} the predicted pairing symmetries of MoTe$_2$ and MoS$_2$ are compared to WS$_2$.  In Sec.~\ref{sec:discussion} we classify the topology of  our predicted gap symmetries and discuss our results in relation to experiment. Finally, Sec.~\ref{sec:conclusion} is devoted to the conclusions.

\section{Theoretical Details}\label{sec:theory}
\subsection{Hamiltonian and Basic Notations}
The Hamiltonian for a general quantum material system with spin and orbital dependent interactions is given by 
\begin{widetext}
\begin{align}
\hat{\mathcal{H}}&=\sum_{\substack{\alpha l\\\beta l^{\prime}}} \int d^{3}r 
\hat{\psi}^{\dagger}_{\alpha l}(r)
h^{0}_{\alpha l,\beta l^{\prime}}(r) 
\hat{\psi}_{\beta l^{\prime}}(r)
+
\frac{1}{2}\sum_{\substack{\alpha\beta\gamma\delta\\ i j k l }}\int \int d^{3}rd^{3}r^{\prime} 
\hat{\psi}^{\dagger}_{\alpha i}(r) \hat{\psi}^{\dagger}_{\beta j}(r^\prime) 
v_{\delta\gamma;\alpha\beta}^{lk;ij}(r,r^{\prime})
\hat{\psi}_{\gamma k}(r^\prime) \hat{\psi}_{\delta l}(r)
\end{align}
\end{widetext}
where the Greek  and Latin letters denote the spin  and orbital degrees of freedom, respectively. Our interaction index notion follows an {\it in$_{r}$in$_{r^{\prime}}$};{\it out$_{r}$out$_{r^{\prime}}$} scheme inline with the diagrammatic representation. For a multiorbital system, the on-site energy of the $l^{th}$ orbital is given by $h^{0}_{\alpha l,\beta l}(r)$ matrix element. If the orbitals on the various atomic sites are close enough for their wave functions to overlap, electrons can hop from one orbital to another. The amplitude of this hopping from orbital $l^{\prime}$ to orbital $l$ is $h^{0}_{\alpha l,\beta l^{\prime}}(r)$. Here, $\mathbf{r}$ is defined over $\mathbb{R}^3$ and the field operators acting on a specific orbital of an atomic site in the crystal, $l$,  can be written as $\hat{\psi}_{l}(\mathbf{r})\equiv\hat{\psi}(\mathbf{r}+\mathbf{R}_{l})$, where $\mathbf{R}_{l}$ is the position of the atom in the unit cell.

The generalized two-particle interaction $v_{\delta\gamma;\alpha\beta}^{lk;ij}(r,r^{\prime})$ is a four-point function that takes the full spin and orbital configuration into account. To gain some intuition into the physical content of $v$ and aid our analysis later, we expand the spin degrees of freedom into the Pauli matrices $\sigma^{x,y,z}$ augmented with the identity $\sigma^{0}$, yielding  
\begin{align}\label{eq:spinchangeofbasis}
v_{\delta\gamma;\alpha\beta}^{lk;ij}(r,r^{\prime})=\sigma^{I}_{\alpha\delta}v_{IJ}^{lk;ij}(r,r^{\prime})\sigma^{J}_{\beta\gamma},
\end{align}
where $I,J\in \{0,x,y,z\}$. Now, the interaction is clearly composed of three distinct classes,  (i) the usual Coulomb interaction, 
\begin{align}
\sigma^{0}_{\alpha\delta}v^{lk;ij}_{00}(r,r^{\prime})\sigma^{0}_{\beta\gamma},
\end{align}
(ii) a spin-spin interaction, 
\begin{align}
\sigma^{n}_{\alpha\delta}v^{lk;ij}_{nm}(r,r^{\prime})\sigma^{m}_{\beta\gamma},
\end{align}
and (iii) a spin-orbit coupling term, 
\begin{align}
\sigma^{n}_{\alpha\delta}v^{lk;ij}_{n0}(r,r^{\prime})\sigma^{0}_{\beta\gamma},
\end{align}
where $n,m\in \{x,y,z\}$. We note that in strongly spin-orbit coupled systems, e.g., heavy fermion systems, the two-particle interaction can be modified to consider $J\cdot J$ coupling rather than the Russell-Saunders $L\cdot S$ coupling.\cite{leighton1959,freeman1967} Moreover, the orbital degrees-of-freedom can also be classified based on the orbital and the point group symmetries of the crystal \cite{bunemann2017coulomb}.

In our numerical calculations below, we will restrict ourselves to considering only local interactions similar to the multiorbital Hubbard model\cite{oles1983antiferromagnetism}. Specifically, we parametrize the bare electron-electron interaction as
\begin{subequations}
\begin{align}\label{eq:LocalInteractions_a}
v_{\sigma\bar{\sigma};\sigma\bar{\sigma}}^{ii;ii} = U, ~~&~~ v_{\sigma\sigma;\sigma\sigma}^{ij;ij} = U^{\prime}, ~~~~  v_{\sigma\bar{\sigma};\sigma\bar{\sigma}}^{ij;ij} = U^{\prime}, \\
v_{\sigma\sigma;\sigma\sigma}^{ij;ji} = J, ~~&~~ v_{\sigma\bar{\sigma};\sigma\bar{\sigma}}^{ij;ji} = J, ~~~~ v_{\sigma\bar{\sigma};\sigma\bar{\sigma}}^{ii;jj} = J^{\prime},\label{eq:LocalInteractions_b}
\end{align}
\end{subequations}
where $U$ is the standard on-site Hubbard term, $U^{\prime}$ characterizes the inter-orbital Coulomb repulsion, $J$ is the so-called Hund's coupling, and $J^{\prime}$ describes spontaneous inter-orbital pair hopping. After rotating into the Pauli basis, the non-zero spin and orbital matrix elements of the bare electron-electron interaction are given in Table~\ref{table:HubbardInteraction}. 

\begin{table}[h]
\Large
\centering
\begin{tabular}{c|c|c|c|c}
$v^{lk;ij}_{IJ}$& $00$ & $xx$ & $yy$ & $zz$  \\
\hline\hline
$ii;ii$ & $\frac{U}{2}$ & $-\frac{U}{2}$ & $-\frac{U}{2}$ & $-\frac{U}{2}$  \\
$ij;ij$ & $U^{\prime}-\frac{J}{2}$ & $-\frac{J}{2}$ & $-\frac{J}{2}$ & $-\frac{J}{2}$  \\
$ij;ji$ & $J-\frac{U^{\prime}}{2}$ & $-\frac{U^{\prime}}{2}$ & $-\frac{U^{\prime}}{2}$ & $-\frac{U^{\prime}}{2}$  \\
$ii;jj$ & $\frac{J^{\prime}}{2}$ & $-\frac{J^{\prime}}{2}$ & $-\frac{J^{\prime}}{2}$ & $-\frac{J^{\prime}}{2}$ 
\end{tabular}
\caption{The non-zero matrix elements of the bare electron-electron interaction parametrized by a set of local spin and orbital dependent Hubbard parameters.}\label{table:HubbardInteraction}
\end{table}

To keep the results and discussion general we define all operators in the {\it imaginary-time} Heisenberg picture,
\begin{align}
\mathcal{O}(z)=U(\tau_0,\tau)\mathcal{O}U(\tau,\tau_0),
\end{align}
with the time arguments, $\tau$ and $\tau_0$, running along the imaginary-axis of the Keldysh contour,  where $\tau_{0}$ is an arbitrary initial time and the time-evolution operator, $U(\tau,\tau_0)$, evolves an operator  $\mathcal{O}$ from   $\tau_{0}$ to $\tau$ along the imaginary-axis. In this picture the operators are explicitly time dependent where as the wave functions are not. This allows us to introduce the time ordering on the contour and Wick's theorem, connecting our results to many-body perturbation theory\cite{stefanucci2013}.

In order to treat the electronic many-body dynamics at finite temperature, we define the time-dependent ensemble average of operator $\mathcal{O}(\tau)$ as 
\begin{align}
\braket{\mathcal{O}(\tau)}=\frac{\text{Tr}\left\{  \mathcal{T} \exp{\left[-\int_{0}^{\beta}  d\bar{\tau} H(\bar{\tau}) \right]    }    \mathcal{O}(\tau)  \right\}    }{\text{Tr}\left\{  \mathcal{T} \exp{\left[-\int_{0}^{\beta}  d\bar{\tau} H(\bar{\tau}) \right]    }    \right\}   },
\end{align}
where $\mathcal{T}$ is the imaginary-time-ordering operator, and  $\braket{\mathcal{O}(\tau)}$ is the overlap between the initial state in thermodynamical equilibrium (for temperature $\beta$) at $\tau_0$ with the time evolved state at $\tau$.

\footnotetext[999]{For an excellent historical overview of the Schwinger Green's function method and Schwinger's personal retrospective on the influence of Green's functions on his work see Ref.~\onlinecite{schweber2005sources} and \onlinecite{schwinger1993greening}.  }

To obtain the exact expression for the effective quasiparticle interactions and the vertex function, along with the various other quantities, we will use the Schwinger functional derivative approach\cite{schwinger1951greenI,schwinger1951greenII,Note999}. To do so, we couple the Hamiltonian to an auxiliary time-dependent electromagnetic field that probes the charge, spin, and orbital degrees of freedom. The coupling between the auxiliary fields and our system is given in a compact form by 
\begin{align}\label{eq:extfield}
\hat{\pi}(\tau_{1})=\int d^2r \pi_{l l^{\prime}}^{I}(1)  \hat{\psi}^{\dagger}_{\alpha l}(1)\sigma_{\alpha\beta}^{I} \hat{\psi}_{\beta l^{\prime}}(1).
\end{align}
Now if we wish to find the infinitesimal change in the ensemble average of a generic, contour-ordered product of operators $\Pi_{i}\mathcal{O}_{i}(\tau_{i})$ with respect to field $\pi_{l l^{\prime}}^{I}(1)$ along the imaginary-time-axis, we arrive at the following identity,
\begin{align}\label{eq:funcderivindentity}
-\frac{\delta}{\delta \pi_{l l^{\prime}}^{I}(1)}&\braket{\mathcal{T}\left\{  \Pi_{i}\mathcal{O}_{i}(\tau_{i})  \right\}   }=\nonumber\\
&\braket{\mathcal{T}\left\{  \Pi_{i}\mathcal{O}_{i}(\tau_{i})      \hat{\psi}^{\dagger}_{\alpha l}(1)\sigma_{\alpha\beta}^{I} \hat{\psi}_{\beta l^{\prime}}(1)  \right\}   }\nonumber\\
-&\braket{\mathcal{T}\left\{  \Pi_{i}\mathcal{O}_{i}(\tau_{i})  \right\}   }\braket{\mathcal{T}\left\{     \hat{\psi}^{\dagger}_{\alpha l}(1)\sigma_{\alpha\beta}^{I} \hat{\psi}_{\beta l^{\prime}}(1)   \right\}   }
\end{align}
In general this identity is valid for equal time and mixed operators, including electronic and bosonic, for more details see Ref. \onlinecite{stefanucci2013}.

\subsection{Relation Between Quasiparticle Interactions and the Vertex Function}

Since the fermionic field operator satisfies the Heisenberg equation of motion
\begin{align}\label{eq:psieom}
\frac{d}{d\tau_{1}}\hat{\psi}_{\alpha n}(1)=\left[  \mathcal{H}, \hat{\psi}_{\alpha n}(1) \right],
\end{align}
we can straightforwardly derive the equation of motion of the single particle Green's function,
\begin{widetext}
\begin{align}\label{eq:G-EOM}
\left(  -\frac{d}{d\tau_{1}} \delta_{l^{\prime}n}\delta_{\alpha\beta}-h^{0}_{\alpha n,\beta l^{\prime}}(1)   \right)G_{\beta l^{\prime},\sigma m}(1,2)
=
\delta(1,2)\delta_{\alpha\sigma}\delta_{nm}
+v^{lk;in}_{\delta\gamma;\xi\alpha}(3,1)G^{(2)}_{\gamma k,\delta l,\xi i,\sigma m}(1,3,3^+,2),
\end{align}
\end{widetext}
where the single- and two-particle Green's functions along the imaginary time axis are given by 
\begin{align}
G_{\beta l^{\prime},\sigma m}(1,2)&=-\braket{         \hat{\psi}_{\beta l^{\prime}}(1)       \hat{\psi}^{\dagger}_{\sigma m}(2)   },\\
G^{(2)}_{\gamma k,\delta l, \xi i,\sigma m}(1,3,3^+,2)&=  \braket{  \hat{\psi}_{\gamma k}(1) \hat{\psi}_{\delta l}(3)     \hat{\psi}^{\dagger}_{\xi i}(3^+)    \hat{\psi}^{\dagger}_{\sigma m}(2)   },
\end{align}
where the superscript $(^+)$ in $ \hat{\psi}^{\dagger}_{\eta j}(3^+)$ denotes this operator should be placed infinitesimally after $\hat{\psi}_{\gamma k}(3) $ when the time ordering operator is applied. The electron creation and annihilation operators were also taken to obey the canonical anti-commutation relations on the contour
\begin{subequations}
\begin{align}
\left\{  \hat{\psi}_{\alpha l}(1) , \hat{\psi}^{\dagger}_{\beta l^{\prime}}(2)  \right\}&=\delta_{\alpha\beta}\delta_{l l^{\prime}}\delta(1-2),\\
\left\{  \hat{\psi}_{\alpha l}(1) , \hat{\psi}_{\beta l^{\prime}}(2)  \right\} &= \left\{  \hat{\psi}^{\dagger}_{\alpha l}(1) , \hat{\psi}^{\dagger}_{\beta l^{\prime}}(2)  \right\}=0,
\end{align}
\end{subequations}
where we have introduced the short hand $\hat{\psi}^{\dagger}_{\beta l^{\prime}}(2) \equiv \hat{\psi}^{\dagger}_{\beta l^{\prime}}(\mathbf{x}_{2},\tau_{2}  )$. For convenience we  use the convention where a repeated index or variable implies a summation or integration, provided the repeated indices are on the same side of the equation. Finally, the full spin and orbital dependent Hedin's equations can be derived, and to be complete we have provided them in Appendix~\ref{Appendix:hedin}.

Since we are interested in describing the interaction between  quasiparticles in an interacting many-body system, and ultimately their pairing, we first relate the two-particle propagator to the scattering vertex to find the effective electron-electron interaction in the interacting system. By using Eq.~(\ref{eq:funcderivindentity}), we can write $G^{(2)}$ in terms of the single-particle propagator as
\begin{widetext}
\begin{align}\label{eq:G2expand}
G^{(2)}_{\gamma k,\delta l,\alpha i,\sigma m}(2,4;3,1)\sigma_{\alpha\delta}^{I}&=G_{\gamma k,\sigma m}(2,1)G_{\delta l,\alpha i}(4,3)\sigma^{I}_{\alpha\delta}
-G_{\gamma k,\mu s}(2,5)\frac{\delta G^{-1}_{\mu s , \nu t}(5,6)}{\delta \pi^{I}_{il}(3,4)}G_{\nu t,\sigma m}(6,1),
\end{align}
\end{widetext}
where we have used the identity, $\frac{\delta G}{\delta \pi}=-G\frac{\delta G^{-1}}{\delta \pi}G$,
to recover the screened vertex function $\frac{\delta G^{-1}}{\delta \pi}$. We further simplify by writing $\frac{\delta G^{-1}}{\delta \pi}$ in terms of the vertex $\frac{\delta \Sigma}{\delta G}$ and iterating the resulting expression to separate the propagators and interaction terms, such as,
\begin{align}
\frac{\delta G^{-1}}{\delta \pi}&=\frac{\delta G^{-1}_{0}}{\delta \pi}+\frac{\delta \Sigma}{\delta G}GG\frac{\delta G^{-1}}{\delta \pi}\\
&=\frac{\delta G^{-1}_{0}}{\delta \pi}-\bar{\Gamma} GG\frac{\delta G^{-1}_{0}}{\delta \pi},\nonumber
\end{align}
with
\begin{align}
\bar{\Gamma}=\left[  \frac{\delta \Sigma}{\delta G}+\frac{\delta \Sigma}{\delta G}GG\frac{\delta \Sigma}{\delta G}+\dots  \right],
\end{align} 
describing a multiple scattering process of two quasiparticles with the vertex. 
Then, by defining the screened, bare, and kernel vertices as
$\Lambda=-\frac{\delta G^{-1}}{\delta \pi}$,
$\Lambda_0=-\frac{\delta G^{-1}_0}{\delta \pi}$,
and $I=-\frac{\delta \Sigma}{\delta G}$, respectively, and inserting $\Lambda$ back into Eq.~(\ref{eq:G2expand}), we find the two-particle Green's function decomposed into three terms,
\begin{widetext}
 \begin{align}
G^{(2)}_{\gamma k,\delta l,\alpha i,\sigma m}(2,4;3,1)=&
G_{\gamma k, \sigma m}(2,1)G_{\delta l,\alpha i}(4,3)\sigma^{I}_{\alpha\delta}-G_{\gamma k,\mu i}(2,3)G_{\nu l,\sigma m}(4,1)\sigma_{\mu\nu}^{I}\\
+&G_{\gamma k,\mu s}(2,5)G_{\nu t,\sigma m}(6,1)\bar{\Gamma}_{\mu s,\eta j,\nu t,\xi n}(5,7;6,8)G_{\xi n , \lambda i}(7,3)G_{\rho l ,\eta j}(4,8)\sigma^{I}_{\lambda \rho}.\nonumber
\end{align}
\end{widetext}
The first two terms of $G^{(2)}$ describe the free propagation of two quasiparticles, where they may flow directly from point to point or exchange positions due to particle indistinguishability. The last term captures the mutual scattering between two quasiparticles due to the effective interaction $\bar{\Gamma}$. To gain more insight into the structure of the effective interaction $\bar{\Gamma}$, we pull out factors of the vertex $\frac{\delta \Sigma}{\delta G}$ from the left and right and re-group the resulting terms, yielding, 
\begin{widetext}
\begin{subequations}
\begin{align}\label{EQ:effective_int}
\bar{\Gamma}_{\mu s,\eta j,\nu t,\xi n}(5,7;6,8)=&I_{\mu s,\eta j,\nu t,\xi k}(5,7;6,8)\\ 
&-I_{\mu s,\theta a,\nu t,\varepsilon b}(5,11;6,12)R_{\theta a,\tau d,\omega c,\varepsilon b}(11,14;13,12)I_{\omega c,\eta j,\tau d,\xi k}(13,7;14,8),
\nonumber\\ \label{EQ:effective_int_ex_response}
R_{\theta a,\tau d,\omega c,\varepsilon b}(11,14;13,12)=&G_{\theta a,\omega c}(11,13)G_{\tau d,\varepsilon b}(14,12)\\ 
&+G_{\theta a,\Delta x}(11,15)G_{\pi y,\varepsilon b}(16,12)\left[-I_{\Delta x,\lambda w,\pi y,\rho m}(15,17;16,18)\right]R_{\lambda w,\tau d,\omega c,\rho m}(17,14;13,18).\nonumber
\end{align}
\end{subequations}
\end{widetext}
$\bar{\Gamma}$ is composed of two classes of quasiparticle interactions, the first an effective {\it direct} interaction and the second an effective {\it exchange} interaction. The direct interaction is constructed from a single $\frac{\delta \Sigma}{\delta G}$ vertex, whereas the exchange part is a generalized infinite ladder sum\cite{fetter2012quantum,mattuck1992guide,abrikosov2012methods}. Here, we have written the generalized ladder sum so as to isolate the system's two-particle exchange response $R$. In general, $R$ is not explicitly solvable, due to the momentum and energy mixing between ladder rungs. However, if the interaction $I$ is a constant or energy independent and separable in momentum\cite{schrieffer2018theory}, e.g. $I(\mathbf{q}-\mathbf{p})=i(\mathbf{q})i^{\prime}(\mathbf{p})$\footnote{For a static potential $I(\mathbf{q}-\mathbf{p})$ this can easily be archived on a finite grid of momenta by diagonalizing with respect to $\mathbf{q}$ and $\mathbf{p}$, yielding 
$I(\mathbf{q}-\mathbf{p})=\sum_{i} V^{\mathbf{q}}_{i} \lambda_{i} V^{\dagger \mathbf{p} }_{i}$, thus converting the ladder sum to a linear algebra problem.}, so as to decouple the rungs, the infinite sum reduces to a geometric series that is readily solvable. By inspection, it is clear that the overall sign of the effective interaction is dictated by the delicate balance between direct and exchange interaction strengths, similar to that found for excitons\cite{onida2002electronic,lane2020interlayer}.  Furthermore, this balance is, in part, governed by the charge, spin, and orbital fluctuations present in the material system.

\subsection{The Effective Pairing Interaction}

\begin{figure*}[t!]
\includegraphics[width=0.98\textwidth]{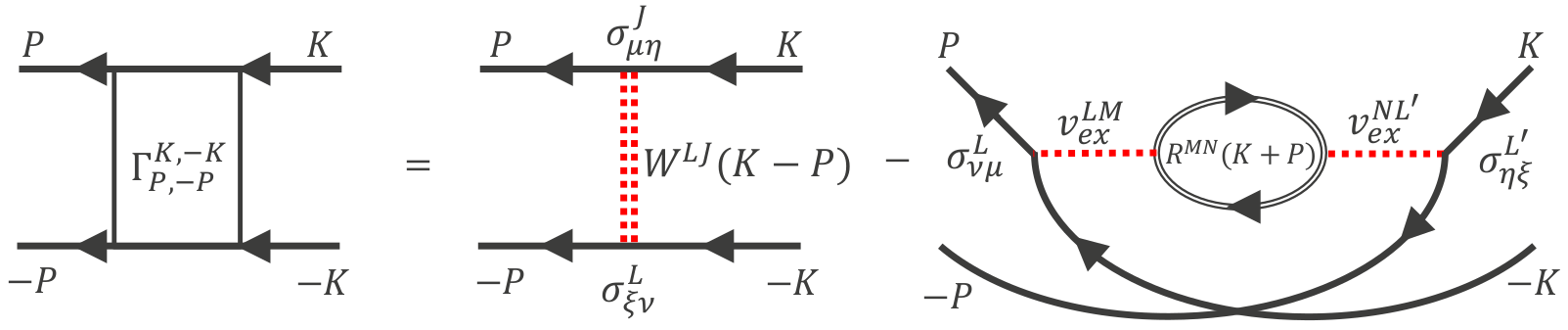}
\caption{(color online) A diagrammatic representation of the effective pairing interaction.} 
\label{fig:EffectiveInteraction}
\end{figure*}

All of the preceding expressions up to this point are exact and fully describe the interaction between two quasiparticles in an interacting many-body system. However, much like Hedin's equations, approximations must be made to make any calculations tractable. To evaluate the vertex $\frac{\delta \Sigma}{\delta G}$ and proceed to characterize the pairing between particles, we invoke the $GW$ approximation for the self energy $\Sigma$. The $GW$ approximation has demonstrated success in providing a reasonable description of the screening environment for 2D materials, yielding excellent predictions of particle-hole pairing (excitons)\cite{qiu2016screening,rasmussen2015computational,cudazzo2016exciton,martin2016interacting}. For more details on the $GW$ approximation, including its physical interpretation, derivation, and consequences, please see Ref.~\onlinecite{hedin1999correlation,aryasetiawan1998gw,onida2002electronic,golze2019gw,martin2016interacting} for a detailed review. Finally, we restrict the generalized two-particle interaction [Eq.~(\ref{eq:spinchangeofbasis})] to be on-site and constant [Eqs.~(\ref{eq:LocalInteractions_a}) and (\ref{eq:LocalInteractions_b})  and Table~\ref{table:HubbardInteraction}]. In the orbital and spin dependent basis the self-energy in the $GW$ approximation is given by 
\begin{align}\label{eq:sigma_gw}
\Sigma_{\mu s,\nu t}(5,6)=&-\sigma_{\mu\gamma}^{J}G_{\gamma k,\alpha a}(5,6)  W^{LJ}_{tk;as}(6,5)\sigma_{\alpha\nu}^{L},
\end{align}
and therefore the vertex $\frac{\delta \Sigma}{\delta G}$ assumes the following form,
\begin{align}
I_{\mu s,\eta j,\nu t,\xi n}(5,7;6,8)=&-\frac{\delta \Sigma_{\mu s,\nu t}(5,6)}{\delta G_{\eta j,\xi n}(7,8)}\\
=&\sigma_{\xi\nu}^{L}W^{LJ}_{tj;ns}(6,5)\sigma_{\mu\eta}^{J}\delta(5,7)\delta(6,8).\nonumber
\end{align}
Inserting the above expressions into Eqs.~(\ref{EQ:effective_int}) and (\ref{EQ:effective_int_ex_response}), converting the spin indices to the Pauli basis using the definitions given in Eqs.~(\ref{eq:spinchangeofbasis}), defining the exchange interaction as
\begin{align}\label{eq:ex_pot}
v_{ex~IJ}^{lk;ij}(r,r^{\prime})=-\frac{1}{4}\sigma^{I}_{\delta\alpha}v_{\delta\gamma;\beta\alpha}^{lk;ji}(r,r^{\prime})\sigma^{J}_{\gamma\beta},
\end{align}
 and applying this effective interaction to two-particles with a zero center-of-mass momentum, we arrive at the effective pairing interaction and corresponding response functions,
\begin{widetext}
\begin{subequations}
\begin{align}\label{eq:pp_effectiveInt}
\bar{\Gamma}^{\mathbf{K} \nu t , -\mathbf{K} \eta j }_{\mathbf{P} \xi n,-\mathbf{P} \mu s}&=\sigma_{\xi\nu}^{L}W^{LJ}_{tj;ns}(\mathbf{K}-\mathbf{P})\sigma_{\mu\eta}^{J}-\sigma^{L}_{\mu\nu}v^{LM}_{ex~ta;sb}R^{MN}_{ad;bc}(\mathbf{K}+\mathbf{P})v_{ex~dj;cn}^{NL^\prime}\sigma^{L^\prime}_{\xi\eta},\\
W^{LJ}_{ak;bn}(\mathbf{K}-\mathbf{P})&= \left[ 1- v^{LM}_{ad;bc}\chi_{0~cf;dg}^{MN}(\mathbf{K}-\mathbf{P})  \right]^{-1}  v_{ak;bn}^{LJ} ,\label{eq:W} \\
R_{ad;bc}^{MN}(\mathbf{K}+\mathbf{P})&= \left[ 1-\chi^{MI}_{0~ay;bx}(\mathbf{K}+\mathbf{P})v^{IK}_{ex~yw;xm}  \right]^{-1}  \chi^{MN}_{0~ad;bc}(\mathbf{K}+\mathbf{P}),\label{eq:R} \\ 
\chi_{0~cf;dg}^{MN}(\mathbf{q})&=\frac{1}{\beta}\sum_{\mathbf{\kappa}}\sigma^{M}_{\alpha^{\prime}\beta^{\prime}}G_{\beta^{\prime} c , \alpha g}(\mathbf{\kappa}+\mathbf{q}) \sigma^{N}_{\alpha\beta}G_{\beta f , \alpha^{\prime} d}(\mathbf{\kappa}),\label{eq:P}
\end{align}
\end{subequations}
\end{widetext}
where the bare interaction was used in the exchange to keep the direct and exchange terms on the same footing. Figure~\ref{fig:EffectiveInteraction} presents a diagrammatic representation of $\bar{\Gamma}$.  Additionally, $\bar{\Gamma}$ is written in $\left|^{in_a,in_b}_{out_a,out_b}  \right.$ notation for clarity and the Matsubara frequencies are suppressed for brevity. Since the Coulomb potential is Hermitian,
\begin{align}
v_{\delta\gamma;\alpha\beta}^{lk;ij} = v_{\alpha\beta;\delta\gamma}^{*ij;lk},
\end{align}
and the polarizability observes the symmetry,
\begin{align}\label{eq:susrule}
\chi_{0~ij;kl}^{\alpha\alpha^{\prime};\beta\beta^{\prime}}(\mathbf{q},\omega) = \chi_{0~kl;ij}^{*\beta\beta^{\prime};\alpha\alpha^{\prime}}(-\mathbf{q},-\omega),
\end{align}
the effective potential $\bar{\Gamma}$ is found to be Hermitian. The nature of $\bar{\Gamma}$ (positive or negative) originates from the spin and charge (orbital) fluctuations in the material, obtained here from the generalized RPA-type expressions for $W$ and $R$\footnote{ We note that due to the multiorbital Hubbard parametrization of the Coulomb interaction, the exchange ladder response $(R)$ takes the form of an RPA-like equation similar to the bubble sum of the direct interaction term $(W)$. In general, however, these two terms are different in nature .}. The emergence of an attractive potential from a nominally repulsive Coulomb interaction follows the spirit of the Kohn-Luttinger mechanism\cite{kohn1965new,luttinger1966new} wherein electron-electron pairing is driven by the intrinsic screening processes present in the many-body electron system. Such an approach has also shed light on the cuprates, the Fe-pnictides, and the doped graphene,\cite{maiti2013superconductivity} where our treatment can be seen as a fully spin- and orbital-dependent generalization of the simple single band case\cite{scalapino1986d,romer2015pairing,romer2020pairing}.

To calculate the effective pairing interaction for a specific material, we assume a non-interacting ground state such that the fully interacting dressed Green's function $(G)$ is replaced by the non-interacting Green's function,
\begin{align}
G_{0~\beta f , \alpha^{\prime} d}(\mathbf{\kappa},i\omega_{n})=\sum_{i}\frac{V_{(\beta f ),i}V^{*}_{(\alpha^{\prime} d),i}}{i\omega_{n}-\varepsilon_{i}},
\end{align}
where $i\omega_{n}$ is the Matsubara frequency and $V_{(\beta f ),i}=\braket{\beta f |i}$  are the matrix elements connecting the orbital-spin and the band spaces found by diagonalizing the Hamiltonian. By introducing the material specific details in this manner, our approach is able to utilize model or {\it ab initio} derived tight-binding Hamiltonians\cite{marzari2012maximally} and projected localized orbitals  derived directly from the Kohn-Sham wave functions\cite{schuler2018charge} allowing for maximum flexibility. Moreover, further correlation and phonon effects maybe straightforwardly accommodated by swapping $G_{0}$ for $G$, and $W$ for $W=W_{ele}+W_{ph}$\cite{hedin1970effects}, respectively. 

Using $G_{0}$, the polarization function [Eq.~(\ref{eq:P})] can be simplified by performing the Matsubara frequency summation and  analytically continuing $i\omega_{n}\rightarrow \omega+i\delta $, for $\delta\rightarrow 0^{+}$. Since we will mainly be concerned with the $\omega\rightarrow 0$ limit, the particle-hole propagator may be recast to produce numerically stable results (see Appendix~\ref{Apendix:lindhard} for details). Additionally, since the pairing potential depends sensitively on the non-interacting polarizability $\chi_{0~ij;kl}^{\alpha\alpha^{\prime};\beta\beta^{\prime}}(\mathbf{q},\omega)$, we enforce the internal symmetry in Eq.~(\ref{eq:susrule}) to remove any spurious numerical differences and guarantee Hermiticity of the pairing potential. 

Finally, the screened direct and exchange response functions are obtained by performing efficient matrix-matrix multiplication and matrix inversion operations on block matrices $\mathbb{D}=v^{LM}_{ad;bc}\chi_{0~cf;dg}^{MN}$ and $\mathbb{F}=\chi^{MI}_{0~ay;bx}v^{IK}_{ex~yw;xm}$. Since, Eqs.~(\ref{eq:W}) and (\ref{eq:R}) take the form of a generalized set of RPA-type equations, extra care must be taken to make sure the various response functions are stable for all temperatures and interaction strengths considered. That is, in the process of solving for $W (R)$ we have introduced the matrix inverse of $\mathbf{1}-\mathbb{D} (\mathbb{F})$. This forces the response functions to be valid if and only if $\mathbf{1}-\mathbb{D} (\mathbb{F})$ is non-singular. Therefore, to check the stability of $W(R)$ for a given set of parameters we diagonalize $\mathbb{D} (\mathbb{F})$ and ensure that the maximum eigenvalue does not exceed $1$. If $\mathbf{1}-\mathbb{D} (\mathbb{F})$ is indeed non-singular, the hierarchy and momentum dependence of the eigenvalue spectrum provides additional insight into the various competing fluctuation modes that may mediate pairing.   

\begin{figure}[t!]
\includegraphics[width=0.99\columnwidth]{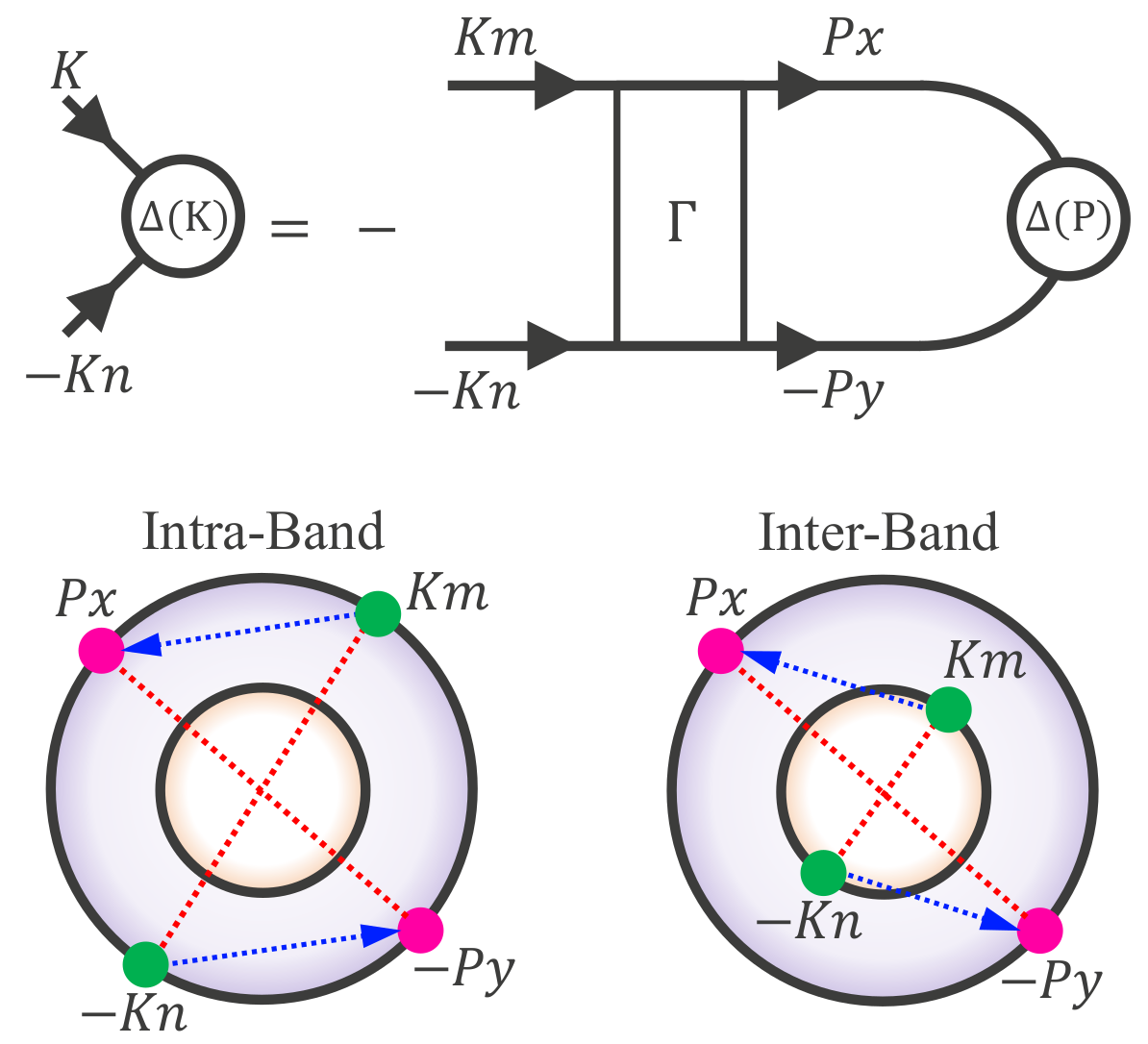}
\caption{(color online) (Top) Diagrammatic representation of the superconducting gap equation in band space. (Bottom) A schematic of the allowed pairing channels: intra-band pairing connects electrons within the same Fermi sheet (black circle), whereas inter-band pairing connects electrons between two different Fermi sheets within the Fermi surface while maintaining a zero center-of-mass momentum. } 
\label{fig:pairingtypes}
\end{figure}

\subsection{The Superconducting Gap Equation}
Following Nozi{\`e}res\cite{nozieres1964theory}, the superconducting gap equation is given by 
\begin{align}\label{eq:gap_orb}
\Delta_{\nu t,\eta j}(\mathbf{K})&=
-\bar{\Gamma}^{\mathbf{K}\nu t,-\mathbf{K}\eta j}_{\mathbf{P}\epsilon b,-\mathbf{P}\theta a}
G_{\omega c,\epsilon b}(\mathbf{P})
G_{\tau d,\theta a}(\mathbf{-P})
\Delta_{\omega c, \tau d}(\mathbf{P})
\end{align}
where $\Delta$ is the momentum, spin, and orbital dependent gap function. Assuming $\Delta$ is frequency independent and $\bar{\Gamma}$ is evaluated at the Fermi level, we can perform the Matsubara frequency sum and convert Eq.~(\ref{eq:gap_orb}) to band space giving,
\begin{align}\label{eq:gap_band}
\Delta_{mn}(\mathbf{K})&=
-\bar{\Gamma}^{\mathbf{K}m,-\mathbf{K}n}_{\mathbf{P}x,-\mathbf{P}y}\lambda^{P}_{xy}
\Delta_{xy}(\mathbf{P}).
\end{align}
Here, we have used the diagonalized-single particle Green's function
\begin{align}
G_{\omega c,\epsilon b}(\mathbf{P})=V^{\mathbf{P}}_{( \omega c ),x} g^{x}(\mathbf{P})V^{\dagger \mathbf{P}}_{x,(\epsilon b)},
\end{align}
and similarly defined the superconducting gap and pairing potential in band space as
\begin{align}\label{eq:gapbasischange}
\Delta_{xy}(\mathbf{P})=V^{\mathbf{P}}_{( \omega c ),x}   \Delta_{\omega c, \tau d}(\mathbf{P})   V^{-\mathbf{P}}_{(\tau d),y}
\end{align}
and
\begin{align}
\bar{\Gamma}^{\mathbf{K}m,-\mathbf{K}n}_{\mathbf{P}x,-\mathbf{P}y}=
V^{\mathbf{K}}_{( \nu t ),m}  V^{-\mathbf{K}}_{( \eta j ),n}
\bar{\Gamma}^{\mathbf{K}\nu t,-\mathbf{K}\eta j}_{\mathbf{P}\epsilon b,-\mathbf{P}\theta a}
V^{\dagger \mathbf{P}}_{( \epsilon b ),x}  V^{\dagger -\mathbf{P}}_{( \theta a ),y},
\end{align}
respectively. Finally, the pairing susceptibility $\lambda^{P}_{xy}$ is defined as,
\begin{align}\label{eq:pairsus}
\lambda^{P}_{xy} = \frac{1-n_{F}^{x}(\mathbf{P})-n_{F}^{y}(-\mathbf{P})}{\Omega_{P}^{x}+\Omega_{-P}^{y}},
\end{align}
where $n_{F}^{x}(\mathbf{P})$ is the Fermi-Dirac function $1/(\exp{(\Omega_{P}^{x}/k_{B}T)}+1)$, and $\Omega_{P}^{x}$ is the quasiparticle energy at momentum $\mathbf{P}$ and band $x$ in the superconducting state.

Since we aim to characterize and predict superconducting order, and their associated symmetries, in real materials, we keep our analysis generic. In contrast to the literature, where it is common to pick a particular pseudospin basis or a specific effective irreducible band symmetry\cite{lindquist2019odd,scaffidi2014pairing,nica2021multiorbital,hu2020pairing,adhikary2020orbital,bandyopadhyay2020superconductivity,ray2019wannier,hsu2017topological},we do not place any such restrictions on the pairing between the various bands (orbitals and spins) configurations present in a given material system\footnote{ We note that interband pairing is typically quite small compared to intraband pairing channels. It is also proposed\cite{moreo2009interband} that when the hybridization among orbitals is strong in a multiband system, both intraband and interband pairings could arise at the Fermi surface obeying Anderson's theorem. In this work, we follow this notion by opening up our approach to this possibility due to the highly hybridized set of bands at the Fermi level in transition metal dichalcogenides. }. This agnostic approach, lets the resulting gap functions $\Delta_{mn}(\mathbf{K})$ inform our symmetry analysis, rather than the other way around, and allows for the emergence of exotic pairing states prevalent in multiorbital systems with spin-orbit coupling and non-symmorphic crystal structures \cite{smidman2017superconductivity,samokhin2020exotic,samokhin2020superconductivity,kim2018beyond,brydon2016pairing}. Figure~\ref{fig:pairingtypes} shows (top panel) a diagrammatic representation of the superconducting gap equations, along with (bottom panel) a schematic of the allowed pairing channels. Intra-band pairing connects electrons within the same Fermi surface (black circles), whereas inter-band pairing connects electrons between two different Fermi surface while maintaining a zero center-of-mass momentum. Furthermore, since incoming (outgoing) electrons can be on different bands, we automatically allow for Fermi surfaces composed of an arbitrary number of degenerate bands.

In general, the self-consistent gap equation [Eq.~(\ref{eq:gap_band})] presents a significant computational challenge. However, to gain insight into the hierarchy of competing superconducting instabilities and their associated pairing symmetry, we consider the solution in the region of $T \approx T_c$. In this case,  $\Delta_{mn}(K) \ll 1$, allowing us to invoke the approximation that the quasiparticle energy spectrum $\Omega_{P}^{x}$ is given by the eigenvalue of the normal-state Hamiltonian, $\varepsilon_{P}^{x}$, thereby linearizing the gap equation. Then by inserting $\Lambda^{\alpha}$ into  Eq.~(\ref{eq:gap_band}) we re-cast the gap equation into a generalized eigenvalue problem,
\begin{align}\label{eq:geneig}
\Delta^{\alpha}_{mn}(\mathbf{K})\Lambda^{\alpha}&=
-\bar{\Gamma}^{\mathbf{K}m,-\mathbf{K}n}_{\mathbf{P}x,-\mathbf{P}y}\lambda^{P}_{xy}
\Delta^{\alpha}_{xy}(\mathbf{P}),
\end{align}
where $\alpha$ enumerates the various superconducting modes and $\Lambda^{\alpha}$ is the corresponding pairing strength. When $\Lambda^{\alpha}=1$ we recover the original gap equation, thus signaling that the normal state is unstable to Cooper pairing. Moreover, it can be shown that the highest eigenvalue has the lowest free-energy in the superconducting  state\cite{scalapino2012common}. Even though the magnitude of $\Delta_{mn}(K)$ no longer has any direct physical meaning, the matrix and nodal structure of the eigengaps still allows us to classify the various pairing symmetries of each superconducting mode. 

Our approach to modeling superconductivity instabilities in quantum material systems is similar to those employed by Hirschfeld {\it et al.}\cite{romer2022leading,hirschfeld2016using,romer2015pairing,roig2022superconductivity} and Scalapino {\it et al.}\cite{graser2009near,altmeyer2016role,scalapino1995case,scalapino1986d}, wherein a realistic multiband treatment is used to evaluate the doping and temperature dependent pairing scenarios within a generalized RPA-type scheme. In the present treatment, we generalize the methodology to allow for strong spin-orbit coupling and non-symmorphic crystal structures, making a realistic examination of topological superconducting candidate materials possible.

\subsection{Symmetry Analysis of the Superconducting Eigengap Solutions}
The analysis of the superconducting eigengap functions maybe carried out from a number of different points of view\cite{wu2015triplet,lu2022chiral,shishidou2021topological,samokhin2019symmetry,hirschfeld2016using,geilhufe2018symmetry,kaba2019group}, such as working within a pseudospin, band, or orbital basis, and so on. Here, we work within the orbital basis, not the band basis, because it allows for a simple description of the symmetries even for intermediate- to strong-coupling regimes\cite{kaba2019group}. We classify the various pairing modes as follows.

A general superconducting gap function is fully antisymmetric under the exchange of the quantum numbers of the pair as dictated by the Pauli exclusion principle, and is diagonal in momentum space assuming translational invariance,
\begin{align}\label{eq:gapantisym}
\Delta^{\alpha}_{\nu t,\eta j}(\mathbf{K}) = - \Delta^{\alpha}_{\eta j,\nu t}(-\mathbf{K}).
\end{align}
To classify the various pairing gap eigenfunctions we express $\Delta_{\nu t,\eta j}$ as a linear combination of spin and orbital basis functions,
\begin{align}
\Delta^{\alpha}_{\nu t,\eta j}(\mathbf{K})=\Delta^{\alpha}_{\tau I}(\mathbf{K})A^{\tau}_{tj}\gamma_{\nu\eta}^{I},
\end{align}
where summation of repeated indices is assumed. The spin part of the pairing $\gamma$ is generally described by the Balian-Werthamer matrices as
\begin{align}
\gamma^I_{\mu\nu}=\left[i\sigma^I\sigma^y\right]_{\mu\nu},
\end{align} 
where $\sigma^I$ is the set of Pauli matrices augmented by the identity matrix $\sigma^{0}$. The three matrices $\gamma^{x,y,z}$ form the symmetric (triplet) part of the spin component of the pairing function, whereas the antisymmetric (singlet) part is represented by the zeroth matrix $\gamma^{0}$.

Similarly, $A^{\tau}$ serves as a basis in orbital space and is defined as a set of $N_{orbital}\times N_{orbital}$ matrices enumerating all intra- and inter-orbital (anti-)symmetric pairing pathways. Specifically, we define the individual intraorbital and interorbital subsets of $A$ as,
{\small
\begin{align}
A^{intra}:\begin{pmatrix}
1 & 0 & 0 & 0 \\
0 & 0 & 0 & 0 \\
0 & 0 & \ddots & \vdots \\
0 & 0 & \cdots & 0 
\end{pmatrix}   ,
\begin{pmatrix}
0 & 0 & 0 & 0 \\
0 & 1 & 0 & 0 \\
0 & 0 & \ddots & \vdots \\
0 & 0 & \cdots & 0 
\end{pmatrix}   ,
\cdots ,
\begin{pmatrix}
0 & 0 & 0 & 0 \\
0 & 0 & 0 & 0 \\
0 & 0 & \ddots & \vdots \\
0 & 0 & \cdots & 1 
\end{pmatrix}   \nonumber
\\
A^{inter}_{sym}:\begin{pmatrix}
0 & 1 & 0 & 0 \\
1 & 0 & 0 & 0 \\
0 & 0 & \ddots & \vdots \\
0 & 0 & \cdots & 0 
\end{pmatrix}   ,
\begin{pmatrix}
0 & 0 & 1 & 0 \\
0 & 0 & 0 & 0 \\
1 & 0 & \ddots & \vdots \\
0 & 0 & \cdots & 0 
\end{pmatrix}   ,
\cdots ,
\begin{pmatrix}
0 & 0 & 0 & 0 \\
0 & 0 & 1 & 0 \\
0 & 1 & \ddots & \vdots \\
0 & 0 & \cdots & 0 
\end{pmatrix}   \nonumber
\\
A^{inter}_{asym}:\begin{pmatrix}
0 & -1 & 0 & 0 \\
1 & 0 & 0 & 0 \\
0 & 0 & \ddots & \vdots \\
0 & 0 & \cdots & 0 
\end{pmatrix}   ,
\begin{pmatrix}
0 & 0 & -1 & 0 \\
0 & 0 & 0 & 0 \\
1 & 0 & \ddots & \vdots \\
0 & 0 & \cdots & 0 
\end{pmatrix}   ,
\cdots ,
\begin{pmatrix}
0 & 0 & 0 & 0 \\
0 & 0 & -1 & 0 \\
0 & 1 & \ddots & \vdots \\
0 & 0 & \cdots & 0 
\end{pmatrix}   \nonumber
\end{align} }
where $A^{intra}$ describes the pairing between electrons occupying the same orbital $j$,
\begin{align}
\hat{\psi}_{\alpha j}(\mathbf{K})\hat{\psi}_{\beta j}(-\mathbf{K}).
\end{align}
In contrast,  $A^{inter}_{sym}$ and $A^{inter}_{asym}$ pick out the symmetric 
\begin{align}
\hat{\psi}_{\alpha j}(\mathbf{K})\hat{\psi}_{\beta i}(-\mathbf{K})+\hat{\psi}_{\alpha i}(\mathbf{K})\hat{\psi}_{\beta j}(-\mathbf{K})
\end{align}
and anti-symmetric 
\begin{align}
\hat{\psi}_{\alpha j}(\mathbf{K})\hat{\psi}_{\beta i}(-\mathbf{K})-\hat{\psi}_{\alpha i}(\mathbf{K})\hat{\psi}_{\beta j}(-\mathbf{K})
\end{align}
pairing between electrons residing on different orbitals $i,j$. \footnote{We note that the total number of $A^{\tau}$ matrices is $N^{2}_{orbital}$, composed of $A^{intra}$, $A^{inter}_{sym}$, $A^{intra}_{asym}$ subsets of sizes $N_{orbital}$, ${N_{orbital} \choose 2}$, and ${N_{orbital} \choose 2}$, respectively.} $\Delta^{\alpha}_{\tau I}$ can be constructed  from our numerical calculations performed in band space by inverting Eq.~(\ref{eq:gapbasischange}) and utilizing the unitarity of $\gamma$ and $A$, 
\begin{align}
\Delta^{\alpha}_{\tau I}(\mathbf{K})=\left[ V^{\dagger\mathbf{K}}_{\nu t,m} \Delta^{\alpha}_{mn}(\mathbf{K})V^{\dagger\mathbf{-K}}_{\eta j,n} \right]
A^{\dagger\tau}_{tj}\gamma_{\nu\eta}^{\dagger I}.
\end{align}
Finally, to gain insight into the dominant pairing gap symmetries it will be useful to take the maximum of $\left| \Delta^{\alpha}_{\tau I}(\mathbf{K})\right|$ over momenta $\mathbf{K}$, to enable a simple comparison of the various matrix elements and construct a table of predicted dominant pairing gap symmetries.

\begin{figure*}[ht!]
\includegraphics[width=0.98\textwidth]{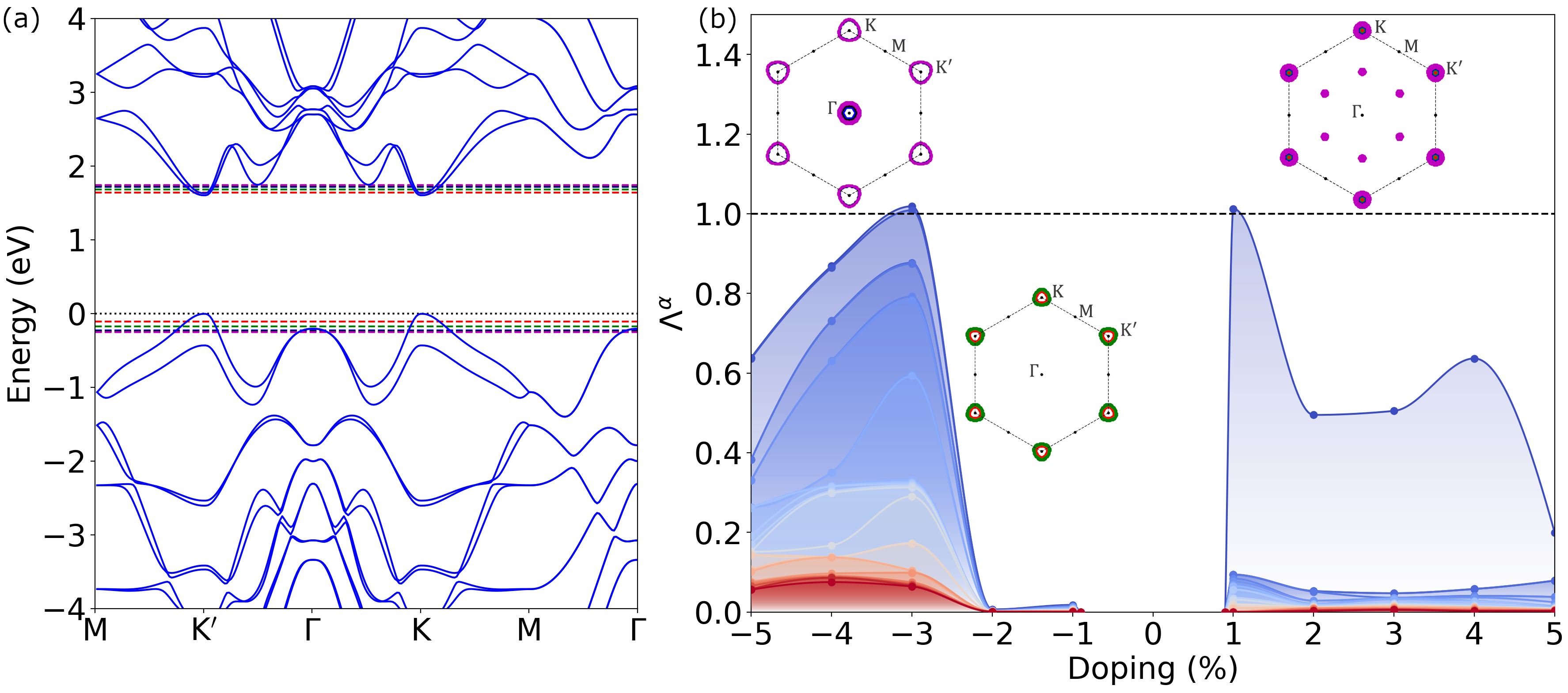}
\caption{(color online) (a) Electronic band dispersion of WS$_2$ (blue solid lines) with the energy cuts at $1\%$, $2\%$, $3\%$, $4\%$, and $5\%$ hole (electron) doping overlaid by dashed red, green, blue, black, violet dashed lines, respectively. (b) Superconducting instabilities $\Lambda^{\alpha}$ as a function of hole (electron) doping with corresponding Fermi surfaces for each unique Fermi surface topology overlaid. The color of the Fermi surface pockets follow those of the energy cuts in (a).} 
\label{fig:BANDS_INSTABILITY}
\end{figure*}

\subsection{Computational Details}
First-principles band structure calculations were carried out within density functional theory framework using the generalized-gradient approximation (GGA) as implemented in the all-electron code WIEN2K~\cite{wien2k19}, which is based on the augmented-plane-wave + local-orbitals(APW+lo) basis set. Spin-orbit coupling was included in the self-consistency cycles. The effective pairing interaction was performed by employing a real-space tight-binding model Hamiltonian, which was obtained by using the wien2wannier interface~\cite{Jan10}. For the various compounds studied, the S-$3p$, Te-$5p$, Mo-$4d$, and W-$5d$ states were included in generating the Wannier functions. The response functions $W(R)$ were evaluated over a $153\times153\times1$ k-mesh. When solving the linearized gap equation we invoke a small energy cutoff $\delta_{c}$ around the Fermi surface, and allow for Cooper pair formation for all electronic states $\varepsilon_{P}^{x} \in \left[ -\delta_{c},\delta_{c} \right]$. The stringency of $\delta_{c}$ was adjusted depending on the size of the various Fermi surface sheets, with values of $\delta_{c}$ ranging between $9.0$ meV and $18.0$ meV. Throughout this work we perform all calculations at 10 K, inline with the typical superconducting transition temperature in the transition-metal dichalcogenides. To limit the number of Hubbard parameters, we assume the Coulomb potential to be rationally invariant, imposing $U^{\prime}=U-2J$, where we have taken $J=J^{\prime}$. We further restrict the value of $J$ to fulfill $J/U=1/6$, which is typical for the transition metals studied. Correlation effects on the sulfur atoms were ignored. Numerically, $U$ is maximized for both electron and hole dopings such that a superconducting instability at 10 K is produced. This procedure, makes $U$ an effective parameter indicating the relative strength of superconductivity in different materials. See Table~\ref{table:HubbardValues} for the values used throughout this work.

\begin{table}[h]
\centering
\begin{tabular}{c|c|c|c|c}
 $x<0$ & $U$ & $U^{\prime}$ & $J$ & $J^{\prime}$  \\
\hline\hline
WS$_2$ & $0.75$ & $0.5$ & $0.125$ & $0.125$ \\ \hline
MoTe$_2$ & $0.3325$ & $0.2216$ & $0.055416$ & $0.055416$ \\ \hline
MoS$_2$ & $0.55$ & $0.366$ & $0.0916$ & $0.0916$ \\ \hline
 $x>0$ & $U$ & $U^{\prime}$ & $J$ & $J^{\prime}$\\
 \hline\hline
 WS$_2$ &  $0.5$ & $0.333$ & $0.083$ & $0.083$ \\ \hline
MoTe$_2$ & $0.3325$ & $0.2216$ & $0.055416$ & $0.055416$ \\ \hline
MoS$_2$ & $0.35$&$0.2333$&$0.05833$&$0.05833$ \\ \hline
\end{tabular}
\caption{The values of the multiorbital Hubbard parameters used in this work.}\label{table:HubbardValues}
\end{table}

\section{Predicted Pairing States in Doped 2H-WS$_2$}\label{sec:pairing}

\subsubsection*{Electronic Structure and Superconducting Instabilities}

Figure~\ref{fig:BANDS_INSTABILITY} (a) shows the Wannier interpolated electronic band structure for monolayer 2H-WS$_2$.  In contrast to the bulk band dispersions, inversion symmetry is broken in the monolayer, producing two inequivalent valleys at the corners of the honeycomb Brillouin zone, labeled by momenta $K$ and $K^\prime$. These extrema points in the valence and conduction bands form a direct band gap. Due to the absence of W $5d_{z^2}$ / S $2p_{z}$ interlayer hybridization\cite{trainer2017inter}, the doubly degenerate bands at $\Gamma$ lie $200$ meV below the Fermi energy. Due to non-zero spin-orbit coupling, especially in the 5$d$ tungston transition metal, a finite spin splitting of $\sim 400$ meV in both conduction and valence valleys is induced. Furthermore, the valleys at $K$ and $K^\prime$ are degenerate and of opposite spin due to time reversal symmetry. 

Since superconductivity sensitively depends both on the pairing glue and number of available carriers, we study a range of finite hole (electron) dopings. As the chemical potential is decreased, hole pockets form at $K$ and $K^\prime$ in the valence band producing a 2D Fermi surface [Fig.~\ref{fig:BANDS_INSTABILITY} (b) insets]. Due to the anisotropy of the band structure surrounding $K(K^\prime)$ the resulting Fermi surface is not circular, but rather a smooth Reuleaux triangle. Passing from 2\% to 3\% hole-doping the Fermi energy cuts additionally through the degenerate bands at $\Gamma$, precipitating a Fermi surface topology change.  The new pocket centered at $\Gamma$ is composed of two concentric  circular Fermi sheets the radii of which increase at slightly different rates with doping. As a consequence of the small spin-orbit induced spin-splitting in the conduction band, two concentric electron pockets are formed at $K(K^\prime)$ for all dopings examined. For 5\% electron doping, several very small additional pockets appear along the radial $\Gamma - K(K^\prime)$ line in the hexagonal Brillouin zone. 

\begin{figure*}[ht!]
\includegraphics[width=0.98\textwidth]{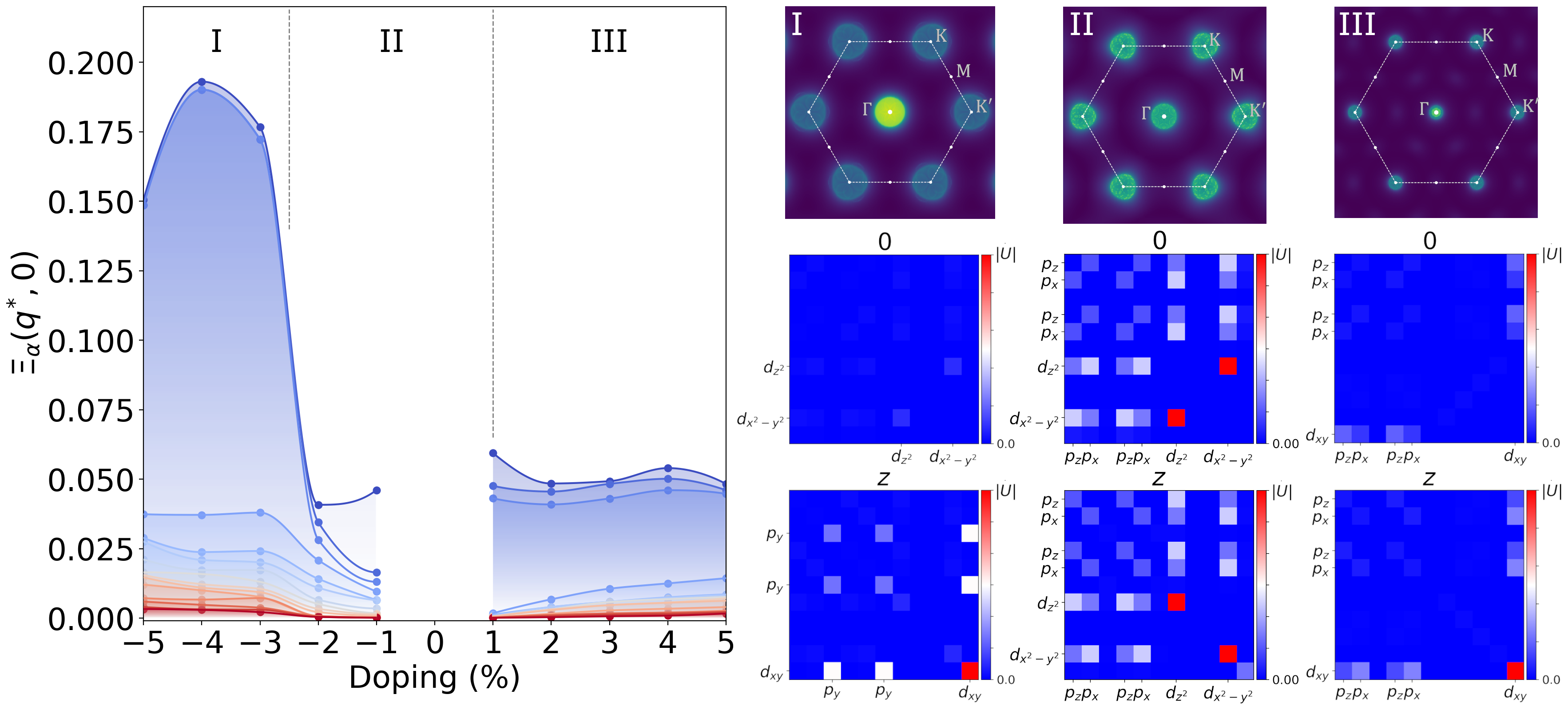}
\caption{(color online) (Left panel) Stoner instabilities $\Xi_{\alpha}(\mathbf{q}^*,0)$ as a function of hole (electron) doping. (Right panel, top) Corresponding full momentum dependence of the leading Stoner instability $\Xi_{\alpha}(\mathbf{q},0)$ for each characteristic doping region, and (right panel, bottom) heat maps of the associated fluctuation character.} 
\label{fig:FLUCTUATIONS}
\end{figure*}

Figure~\ref{fig:BANDS_INSTABILITY} (b) presents the leading 20 superconducting instabilities (blue to red shaded regions) obtained by solving the generalized eigenvalue problem in Eq.~(\ref{eq:geneig}) for various hole (electron) dopings at 10 K. The associated Fermi surfaces are given as insets. Starting on the hole-doped side, we find $x= -0.01$ and $-0.02$ concentration of hole carries to exhibit a very weak pairing strength of $\sim 10^{-2}$. The leading instability is non-degenerate, with the other subleading $\Lambda$ just $2.0\times 10^{-3}$ below. Passing through the Fermi surface topological transition the leading pairing strength increases by $60$ times to $1.039$ signaling an instability of the ground state to superconductivity. Here, the leading instability is nearly degenerate, with a marginal splitting of $0.01$. As the hole doping is increased, the leading, and subleading, instabilities decrease roughly linearly. 

Turning to the electron doped instabilities, by inspection we find the leading instability to be non-degenerate and separated by an order of magnitude from the subleading pairing strengths for all dopings considered. An instability to a superconducting ground state is predicted for 1\% electron doping, followed by a non-monotonic decrease in the pairing strength with increased doping. Finally, for $x=0.05$ the pairing strength significantly weakens to $0.2$, which is concomitant with the addition of the pockets along the $\Gamma - K(K^\prime)$ path.

\subsubsection*{Charge and Spin Fluctuations}
The non-monotonic evolution of the superconducting pairing instabilities with doping follows concomitantly the changes in Fermi surface topology, as is suggested by the insets in Fig.~\ref{fig:BANDS_INSTABILITY} (b). Since the effective potential $\bar{\Gamma}$ is constructed from generalized RPA-type response functions, there is indeed an intimate connection between Fermiology and the strength of the pairing potential. Specifically, information of the Fermi surface is encoded as peaks in the polarizability, where these resonances appear at special momenta $\mathbf{q^*}$ that facilitate Fermi surface nesting. Moreover, the inclusion of interactions within this RPA-type response further enhances existing features in $\chi_{0}$ as the generalized Stoner denominator $1-\mathbb{D}(\mathbb{F})$ approaches zero. Therefore, by analyzing the dominant fluctuations in the system, we can gain some physical insight into the doping dependence of the pairing instabilities.

To extract the various fluctuation modes and their associated spin-orbital character, we recognize that the Stoner instabilities can be made transparent by diagonalizing the complicated kernel present in Eq.~(\ref{eq:R}) and (\ref{eq:W}). That is, for $\mathbb{F}=\chi_{0}v_{ex}$ in $R$,
\begin{align}
\mathbb{F}&=U(\mathbf{q},\omega) \Xi(\mathbf{q},\omega)U^{-1}(\mathbf{q},\omega),
\end{align}
where $\Xi$ is a diagonal matrix, and $U$ is unitary, then
\begin{align}
R^{MN}&(\mathbf{q},\omega)= \\
& U_{M\alpha}(\mathbf{q},\omega)\left[ 1-\Xi_{\alpha}(\mathbf{q},\omega) \right]^{-1}U_{\alpha K}^{-1}(\mathbf{q},\omega)  \chi^{KN}_{0~}(\mathbf{q},\omega),\nonumber
\end{align}
where $\alpha$ enumerates the fluctuation eigenmodes. Now, as $\Xi_{\alpha}(\mathbf{q},\omega=0)$ approaches $1$ the ground state becomes unstable to an ordered phase. Additionally, the momenta producing the maximum $\Xi_{\alpha}$ for a given $\alpha$ sheet, is the propagating vector $\mathbf{q}^{*}$ of the emerging Stoner instability. The character of this instability may then be obtained by analyzing the associated eigenvectors, $U$. The fluctuation eigenmodes for $\mathbb{F}$ and $\mathbb{D}$ were found to be the same. Therefore, without loss of generality, we only discuss those from $\mathbb{F}$ in the following discussion.

Figure~\ref{fig:FLUCTUATIONS} (left panel) shows the first 15 fluctuation eigenmodes $\Xi_{\alpha}(\mathbf{q}^{*})$ as a function of doping. To facilitate discussion, the doping range has been subdivided into three characteristic regions based on the Fermi surface topology, labeled in Fig.~\ref{fig:FLUCTUATIONS} as I, II, and III. The full momentum dependence of the leading Stoner instability $\Xi_{\alpha}(\mathbf{q},0)$ and associated character in each region is presented columnwise in the right hand panel.

\begin{figure*}[ht!]
\includegraphics[width=0.98\textwidth]{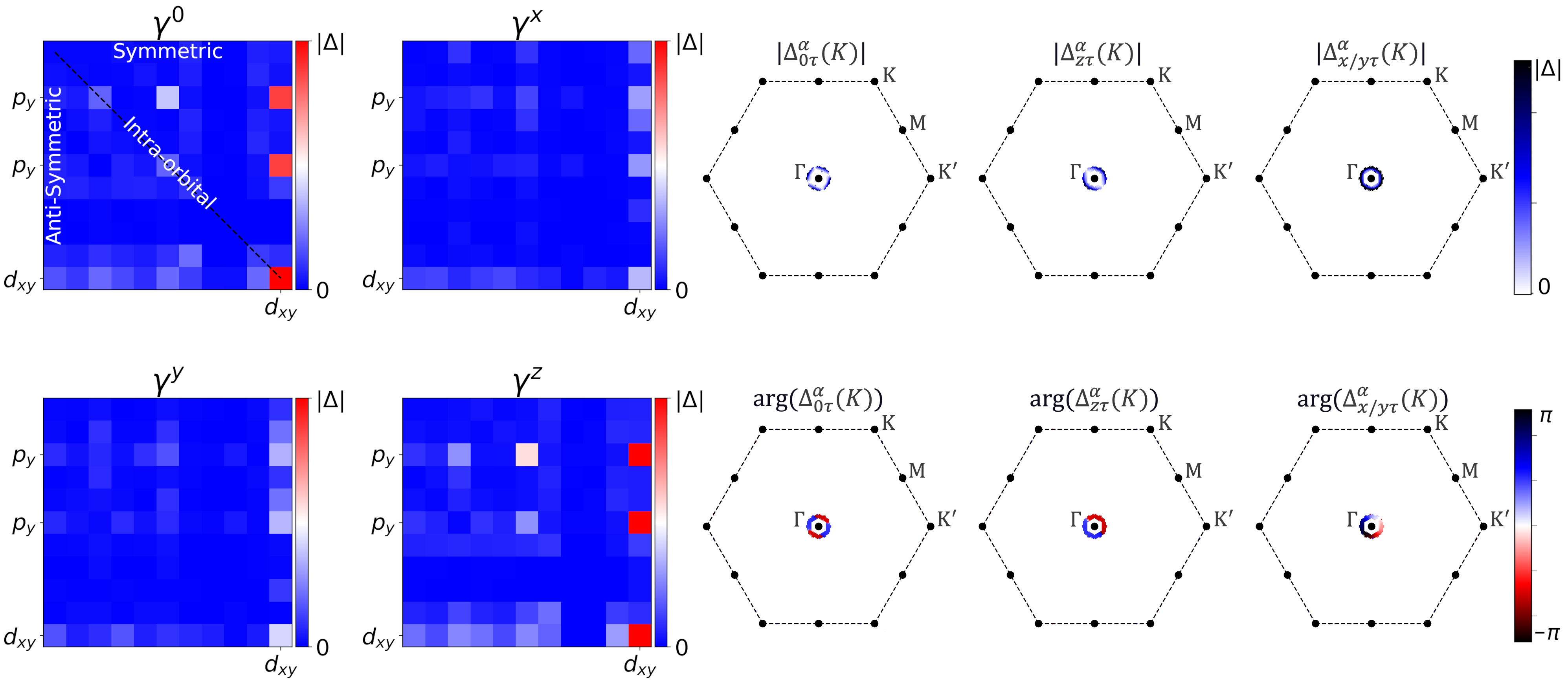}
\caption{(color online) (Left panel) Heat map of the various superconducting eigen gap matrix elements given by the max of $\left| \Delta^{\alpha}_{\tau I}(\mathbf{K})\right|$ over momenta $\mathbf{K}$ for the leading pairing mode in WS$_2$ with 3\% hole doping at 10 K. The orbitals contributing to the dominant pairing channels are indicated. (Right panel) The variation of the superconducting eigen gap amplitude and phase over the Fermi surface for the two dominant $\gamma^{0},\gamma^{z}$ and subdominant  $\gamma^{x}(\gamma^{y})$ channels present in the left panel (red cells). The momentum dependence of the gap function is the same for each orbital combination.} 
\label{fig:valence_gap_symmetry}
\end{figure*}

For dilute hole doping, $\mathbf{q}^{*}$ facilitates both intra- and inter-pocket nesting, producing plateaus of similar weight around $\Gamma$ and $K(K^{\prime})$, except for a slightly larger peak at $\mathbf{q}^{*}=0$ due to perfect nesting between the degenerate bands. The heat maps in the bottom two panels of column II in Fig.~\ref{fig:FLUCTUATIONS} (right panel), displays the relative contribution each spin and orbital plays in the given mode. For II, this mode is  predominantly composed of charge and longitudinal spin sectors, displaying strong coupling between tungsten $5d_{x^2-y^2}$ $(5d_{xy})$ and sulfur $3p$ orbitals. Crossing through the Fermi surface topological transition, between 2\% to 3\% hole carriers, there is a sharp rise in the leading Stoner instability, reaching a maximum of nearly 0.2 for $x=-0.04$. Due to the perfect $(\mathbf{q}=0)$ nesting between the additional doubly degenerate Fermi pocket at $\Gamma$, $\mathbf{q}^{*}\approx 0$ scattering dominates the instability spectrum. Interestingly, the leading instability is comprised of nearly pure longitudinal spin fluctuations driven by W-$5d_{xy}$ and S-$3p_{y}$ orbitals. As hole doping is increased beyond 4\%, the Stoner instabilities decrease. This effect stems from the strong spin-orbit coupling induced band splitting near $\Gamma$, which breaks the band degeneracy, and thus suppresses $\mathbf{q}^{*}=0$ nesting.

Region III covers the full electron doping range considered.  Here, the two concentric Fermi surfaces at $K(K^{\prime})$ exhibit a similar fluctuation map as region II, except enhanced instabilities near $\Gamma$ due to the additional pocket. Despite the value of $U$ being $1.5\times$ smaller on the electron doped side, the Stoner instabilities are similar in strength to those in region II. Lastly, these fluctuations are mainly constructed from the W-$5d_{xy}$ states in the longitudinal spin sector, with a faint admixture of sulfur $3p_{x}$ orbitals.

By comparing the fluctuation eigenmodes and superconducting instabilities, we can identify the key mechanism driving doping dependence. The significant increase in pairing strength between 2\% to 3\% hole-doping can be explained by enhanced nesting between the double degenerate bands at $\Gamma$. Moreover, the non-monotonic evolution of the electron doped superconducting instabilities follows the change in the fluctuation strengths. However, we should stress the mapping between $\Xi_{\alpha}(\mathbf{q}^{*})$ and $\Lambda_{\alpha}$ is not one to one. If we consider $x<-0.03$, $\Lambda_{\alpha}$ decreases, while $\Xi_{\alpha}(\mathbf{q}^{*})$ still increases with doping. Moreover, all electron doped superconducting instabilities are very strong, even with a smaller $U$. This issue directly emphasizes the subtle point that the effective potential is the {\it difference} between $W(\mathbf{K-P})$ and $v_{ex}R(\mathbf{K+P})v_{ex}$, making the existence of large fluctuations a necessary rather than sufficient condition for strong pairing.

\begin{figure*}[ht!]
\includegraphics[width=0.98\textwidth]{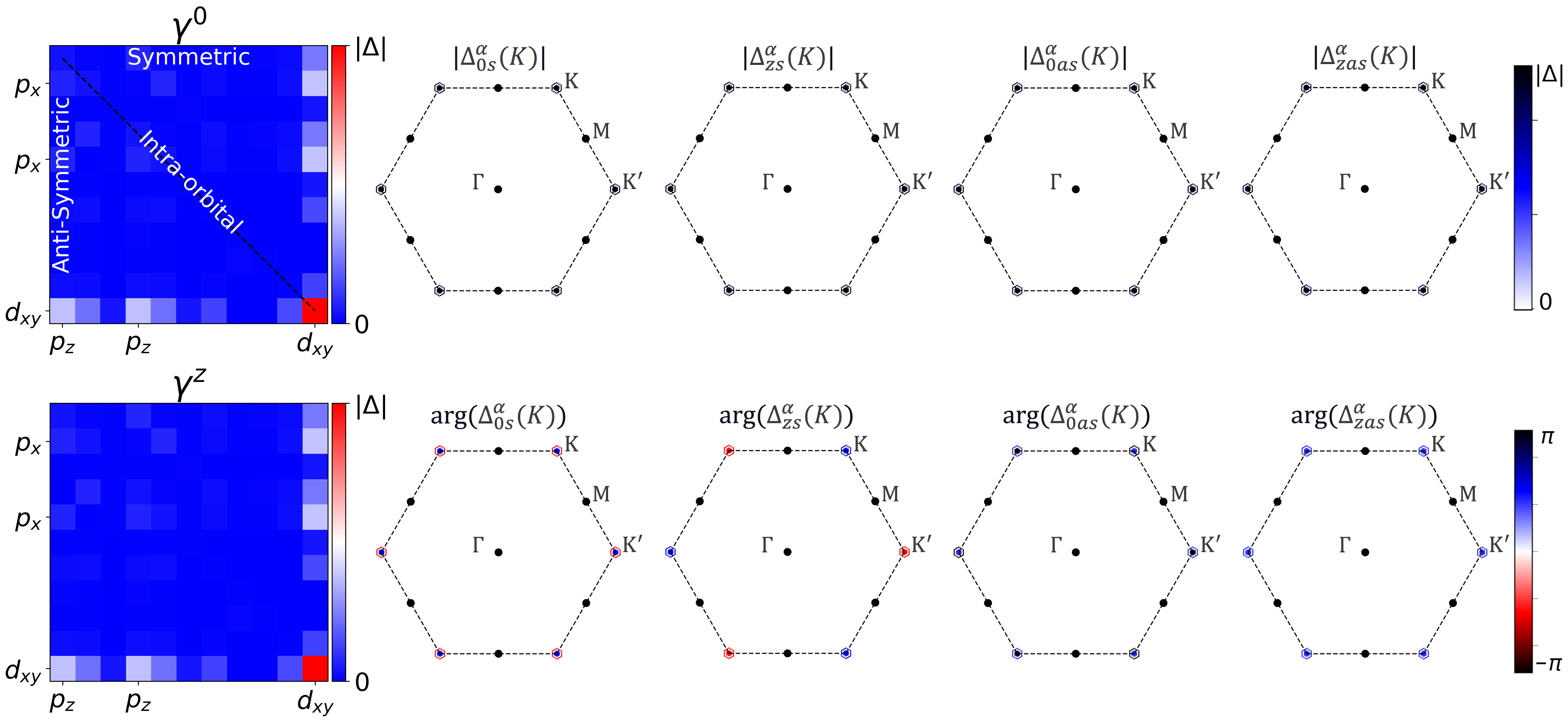}
\caption{(color online) (Left panel) Heat map of the various superconducting eigen gap matrix elements given by the max of $\left| \Delta^{\alpha}_{\tau I}(\mathbf{K})\right|$ over momenta $\mathbf{K}$ for the leading pairing mode in WS$_2$ with 1\% electron doping at 10 K. The matrix elements for $\gamma^{x}$ and $\gamma^{y}$ have been omitted for brevity since the various values were at least an order of magnitude smaller than those in the present panels. The orbitals contributing to the dominant pairing channels are indicated. (Right panel) The variation of the superconducting eigen gap amplitude and phase over the Fermi surface for the dominant $A^{intra}$ and $A^{inter}_{sym}$ symmetric and $A^{inter}_{asym}$ anti-symmetric components  in the $\gamma^{0}$ and $\gamma^{z}$ spin channels present in the left panel (red cells).} 
\label{fig:conduction_gap_symmetry}
\end{figure*}

\subsubsection*{Gap Symmetries}

The gap symmetry plays one of the most important roles in determining whether a superconductor is topologically trivial or not. Similar to the parity analysis of single electron states in topological band theory\cite{bansil2016colloquium}, the variation and sign change of the superconducting gap along the Fermi surface is a key decisive ingredient in classifying the topological nature of the superconducting ground state. Despite evidence for odd-parity pairing states being reported in Sr$_2$RuO$_4$\cite{maeno1994superconductivity,maeno2011evaluation}  and UPt$_3$\cite{tou1998nonunitary,stewart1984possibility}, no clear-cut bridge between odd-pairing symmetries and real topological superconductors has been established.

To identify possible intrinsic topological superconductivity in 2D TMDCs, we analyze and classify the various symmetries and pairing channels of the eigengap functions of the predicted superconducting ground states. For brevity, we will present a detailed analysis of the leading 3\% hole and 1\% electron doped pairing modes in 2H-WS$_2$, giving the associated eigengap symmetries for the remaining instabilities in Table~\ref{table:GapSymsWS2}.

Figure~\ref{fig:valence_gap_symmetry} (left panel) shows a heat map of the maximum of the superconducting eigengap function $\left| \Delta^{\alpha}_{\tau I}(\mathbf{K})\right|$ over momenta $\mathbf{K}$ for the various spin and orbital matrix elements. Despite the lack of inversion symmetry, strong spin-orbit coupling, and significant orbital mixing in the valance band, most  components of $\left| \Delta^{\alpha}_{\tau I}\right|$ are vanishingly small, except for three dominant components. These components arise in the spin singlet $(\gamma^{0})$ and spin triplet $(\gamma^{z})$ sectors, with orbital pairing matrix elements exhibiting $A^{intra}$ and $A^{inter}_{sym}$ pairing composed of W-$5d_{xy}$ and S-3$p_{y}$/W-5$d_{xy}$ orbitals, respectively. Further subdominant modes are identified in the $\gamma^{x}$ $(\gamma^{y})$ spin-triplet component with the same orbital pairing character.

Figure~\ref{fig:valence_gap_symmetry} (right panel) displays the amplitude and phase of $\Delta^{\alpha}_{\tau I}(\mathbf{K})$ in the Brillouin zone, for the dominant and sub-dominant pairing pathways. For $\gamma^{0}$, the superconducting eigengap amplitude displays a finite nodal four-fold symmetry and a phase that alternates between $\pi$ and $-\pi$ as one traverses the $\Gamma$-centered hole packet Fermi surface. This structure in momentum space is indicative of a $d$-wave superconducting gap. Additionally, since the nodal lines are slightly tilted off the horizontal (vertical) axis, we can also conclude this $d$-wave state is a linear combination of both $d_{x^2-y^2}$ and $d_{xy}$ form factors coupled by a real constant governing the tilt angle. Similarly, $\Delta^{\alpha}_{\tau z}(\mathbf{K})$ displays a slightly of off-axis two-fold symmetric eigengap amplitude and alternating phase along the $\Gamma$-centered Fermi surface implying a $p$-wave superconducting state.

Interestingly, the sub-dominant eigengap function in the $\gamma^{x}$ $(\gamma^{y})$ spin-triplet sector fully gaps out the $\Gamma$-centered Fermi surface similar to an $s$-wave state. However, unlike a constant phase expected for a uniform pairing gap, its phase winds around the Fermi surface crossing zero once. This implies a chiral superconducting order of the form $p_x \pm i p_y$. Combining all the eigengap function terms, the predicted gap-function symmetry can be written as
\begin{align}
&\Delta^{\alpha}=\\
&\left[ d \gamma^{0} + (p \pm i p)(\gamma^{x}+\gamma^{y})+p \gamma^{z}\right](A^{d_{xy}}+A^{d_{xy}/p_{y}}_{sym}).\nonumber
\end{align}
We note our numerical scheme intrinsically predicts $\Delta^{\alpha}_{\tau I}(\mathbf{K})$ to be fully anti-symmetric, thus, satisfying the Pauli exclusion principle [Eq.~(\ref{eq:gapantisym})] without needing to force any constraints. Finally, the presence of this pairing symmetry in the leading superconducting instability, suggests a possible topological superconducting ground state in 3\% hole-doped 2H-WS$_2$. 

\begin{figure}[hb!]
\includegraphics[width=0.98\columnwidth]{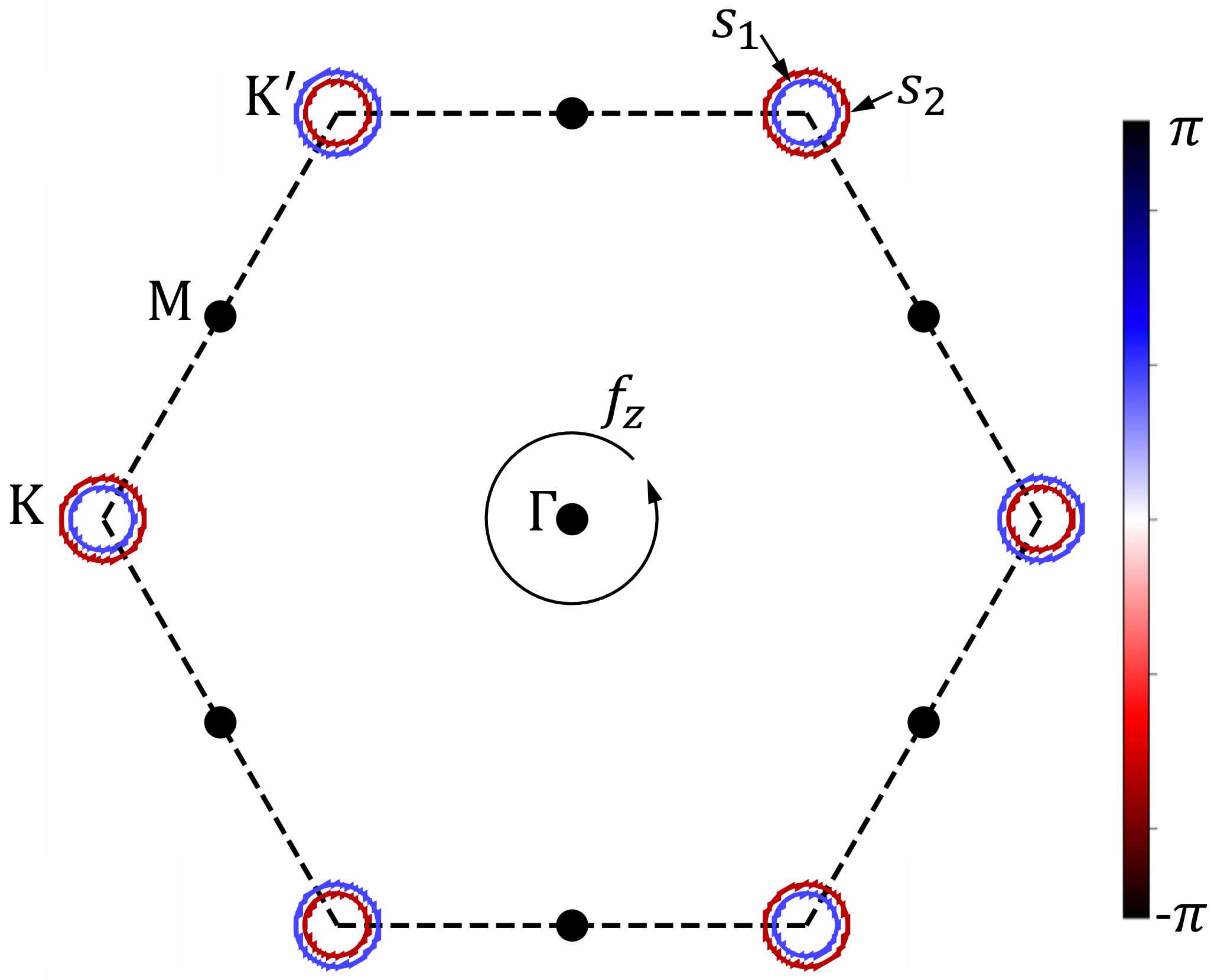}
\caption{(color online) The phase of the superconducting eigengap along the Fermi surface for 5\% electron doped WS$2$ in the $A^{d_{xy}/p_z}_{asym}\gamma^{0}$ sector. The full gap symmetry $(s_1 - s_2)f_z$ is composed of two pocket centered s-wave form factors ($s_1$ and $s_2$) with opposing phases and an alternating phase between $K$ and $K^\prime$ pockets as seen from the zone center following an $f$-wave periodicity.} 
\label{fig:gap_symmetry_diagram}
\end{figure}

Figure~\ref{fig:conduction_gap_symmetry} shows the same as Fig.~\ref{fig:valence_gap_symmetry} except for 2H-WS$_2$ under 1\% electron doping. Interestingly, only $\gamma^{0}$ and $\gamma^{z}$ display non-negligible matrix elements. In particular, a single dominant $A^{intra}$ W-$5d_{xy}$  pairing channel is exhibited, with additional sub-dominant $A^{inter}_{sym}$ $(A^{inter}_{asym})$ S-3$p_{x}$/W-5$d_{xy}$  (S-3$p_{z}$/W-5$d_{xy}$) pairing. Since the electron packet Fermi surfaces are centered at $K$ and $K^{\prime}$, and not at $\Gamma$, we classify the momentum dependence of $\Delta^{\alpha}_{\tau I}(\mathbf{K})$ in two parts: (i) by the variation of the gap relative to the center of each $K(K^{\prime})$ Fermi pocket, and (ii) relative to the zone center to capture the relative phase between pockets. To clearly communicate the various gap symmetries we adopt the notation $(t_{1}\pm t^{\prime}_{2})t^{\prime\prime}_z$, where $t,t^{\prime},$ and $t^{\prime\prime}$ specify the type of superconducting gap, e.g. $s$, $p$, $d$, $f$, $g$, and so on, the subscripts denote whether it is the inner (outer) Fermi surface or zone centered (z), and $\pm$ indicates the relative phase between Fermi sheets. For example, Fig.~\ref{fig:gap_symmetry_diagram}  shows the eigengap function for an anti-symmetric inter-orbital pairing in the spin-singlet sector displaying a constant gap amplitude and phase across both inner and outer Fermi sheets encirculing $K(K^\prime)$, denoted by $s_1$ and $s_2$, respectively. Additionally, a phase difference of $\pi$ is found between Fermi surfaces. Relative to the zone center, the phase alternates by a factor of $\pi$ as one traverses around the Brillouin zone edge. Hence, we can classify this eigengap as $(s_1-s_2)f_z$, an inhomogeneous  $s$-wave superconductor. Similarly, for Fig.~\ref{fig:conduction_gap_symmetry} the $A^{intra/inter}_{sym}\gamma^{0}$, $A^{intra/inter}_{sym}\gamma^{z}$, $A^{inter}_{asym}\gamma^{0}$, and $A^{inter}_{asym}\gamma^{z}$ components have symmetries $(s_1-s_2)s_z$, $(s_1+s_2)f_z$, $(s_1-s_2)f_z$, $(s_1+s_2)s_z$, respectively. The above analysis, then predicts the gap symmetry for the 1\% electron doped WS$_2$ to be,  
\begin{widetext}
\begin{align}
\Delta^{\alpha}= \left( (s_1-s_2)s_z \gamma^{0} + (s_1+s_2)f_z \gamma^{z} \right)
 \left( A^{d_{xy}}+ A^{d_{xy}/p_{x}}_{sym} \right)  
 +
\left( (s_1-s_2)f_z \gamma^{0} + (s_1+s_2)s_z \gamma^{z} \right)
 A^{d_{xy}/p_{z}}_{asym}.
\end{align}
\end{widetext}
We would like to point our that our agnostic approach to analyzing the superconducting gap symmetry is well suited and scales well for real material systems with multiple bands and possible mixed parity. 

\section{Comparison of Predicted Pairing States in Doped WS$_2$,WS$_2$, MoTe$_2$, and 2H-MoS$_2$}\label{sec:compare_pairing}

Since, the effective same-spin pairing interactions can be enhanced by spin-flip processes in the polarizability, the intrinsic spin-orbital coupling of a material strength is a key quantity. Therefore, to design materials with topological superconductivity it is important to have fine control over the spin-orbi- coupling strength. Fortunately, owning to the tunability of the TMDC class of materials, we are able to adjust the intrinsic spin-orbit coupling strength of the pristine compound by substituting the transition metal (chalcogenides) away for different atomic species in the same column of the periodic table. 

\begin{figure*}[ht!]
\includegraphics[width=0.98\textwidth]{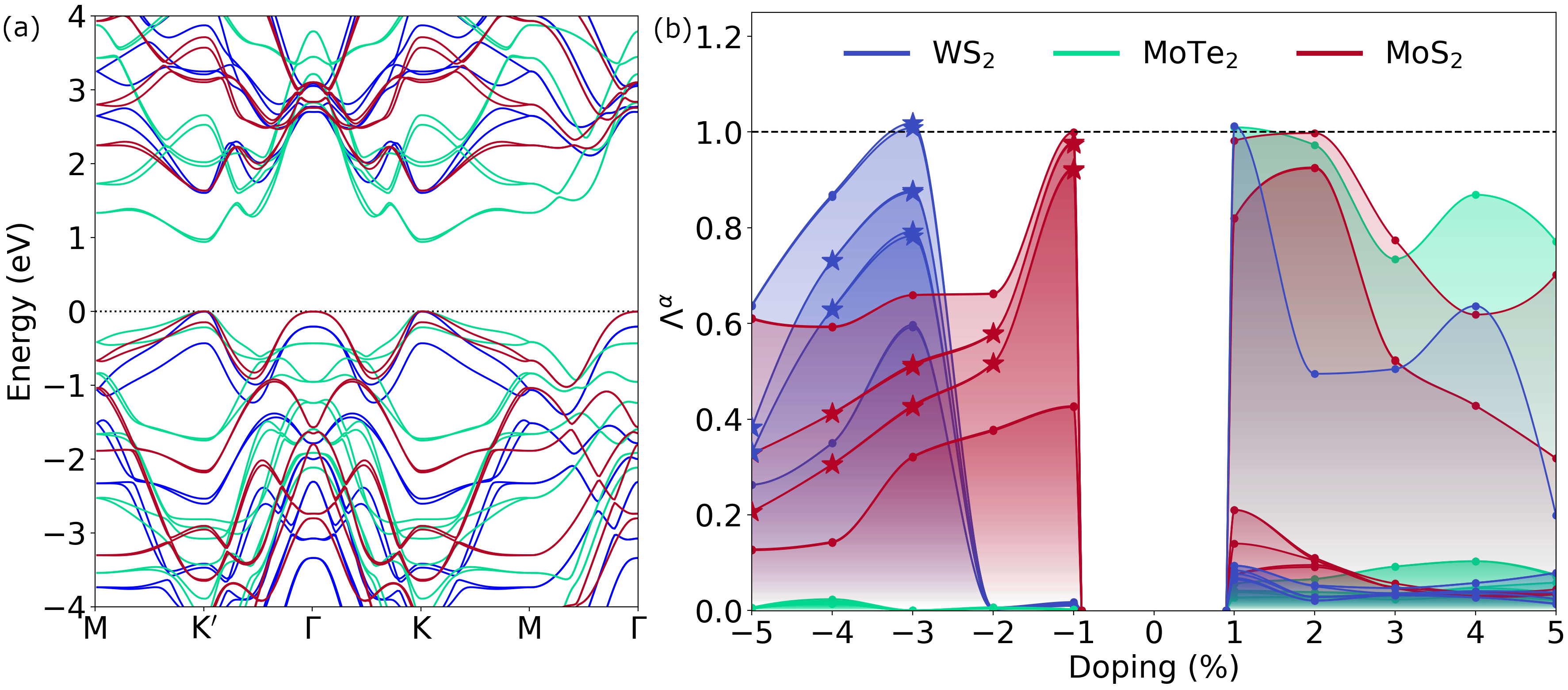}
\caption{(color online) Comparison of the  (a) electronic band dispersion and (b) leading eight superconducting instabilities $\Lambda^{\alpha}$ for WS$_2$, MoTe$_2$, and MoS$_2$ at 10 K. Chiral superconducting modes are indicated by star symbols.} 
\label{fig:comparison_instability}
\end{figure*}

Figure~\ref{fig:comparison_instability} (a) directly compares the Wannier interpolated electronic band structure of monolayer WS$_2$, MoTe$_2$, and MoS$_2$ in the 2H phase. When tungsten is substituted for molybdenum, the large spin splitting of ~ 433 meV reduces by a factor of 3 to 148 meV due to the reduced spin-orbit coupling strength.  Concomitantly, the valence bands at $\Gamma$ raises in energy stopping within 3.7 meV of the Fermi energy. Interestingly, the conduction band minima at $K$ and $K^{\prime}$ are minimally affected, evincing only a slight flattening of the bands. When sulfur is replaced with the sightly heavier tellurium, there is a marginal increase of 69 meV in the spin splitting at $K (K^\prime)$. Moreover, the valence bands at $\Gamma$ drops in energy burying its self to 433 meV binding energy. Surprisingly, the band gap has shrunk by almost a factor of 2, exhibiting a small spin splitting of 34 meV in the conduction band minima.

Figure~\ref{fig:comparison_instability} (b) presents the leading eight superconducting instabilities in WS$_2$, MoTe$_2$, and MoS$_2$ (blue, green, red shaded regions, respectively) obtained by solving the generalized eigenvalue problem in Eq.~(\ref{eq:geneig}) for various hole (electron) dopings at 10 K. Chiral superconducting modes are indicated by star symbols. Firstly, we note that there is no significant rise in the pairing strength of MoTe$_2$ within the hole doping range studied, leaving all pairing strengths at or below 0.024. This is due to the doubly degenerate valence bands at $\Gamma$ being buried in energy, requiring a substantial hole doping to activate these bands. 

When tungsten is replaced with molybdenum, there is a marked change in the pairing strength eigenvalue spectrum. Specifically, the pairing response on the hole doped side appears to be stronger since the effective Hubbard $U$ needed to make the system unstable to Cooper pairing has reduced from  $0.55$ eV to $0.35$ eV. Furthermore, there is now substantial competition between pairing modes. In WS$_2$ the leading six modes span values 0.78 - 1.0, whereas in MoS$_2$ the first six instabilities are within 0.08 of the critical pairing strength.  Although the leading instability is not chiral at $x=-0.01$, the competing pairing modes are strong chiral states, suggesting that slight environmental perturbations may tip the scales, producing a strong topological superconducting ground state. Furthermore, we attribute the lack of leading chiral modes in MoS$_2$ to the reduced spin-orbital coupling strength as compared to WS$_2$, thereby illustrating the material-specific nature of the possible superconducting states.

For electron-doped TMDCs considered we find significant similarities. The evolution of pairing strength with electron doping in WS$_2$ and MoTe$_2$ is quite comparable. Both systems display a leading instability at $x=0.01$ and then monotonically decay, precipitating another peak at 4\% electron doping. On the other hand MoS$_2$ exhibits characteristically different features.  In this case, the leading instability is nearly flat between $0.01<x<0.02$, with the peak at 2\% electron doping. A second pairing instability is 0.92 below, seemingly due to the weakened spin-orbit coupling, yielding nearly degenerate conduction bands. Curiously, all pairing symmetries are similar between the three materials (see Tables~\ref{table:GapSymsWS2} - \ref{table:GapSymsMoTe2_2}).

\section{Discussion}
\label{sec:discussion}

To date, there have been a number of experiments performed within the TMDC class of materials looking for and examining the superconducting phase\cite{liu2021discovery,manzeli20172d,shi2015superconductivity,biscaras2015onset,costanzo2016gate,costanzo2018tunnelling,lu2015evidence,jo2015electrostatically,ye2012superconducting,taniguchi2012electric,li2021recent}  . In particular, MoS$_2$\cite{taniguchi2012electric,ye2012superconducting,shi2015superconductivity} has been the most studied, with a few reports on WS$_2$\cite{biscaras2015onset,costanzo2016gate,costanzo2018tunnelling,lu2015evidence,shi2015superconductivity,jo2015electrostatically} and MoTe$_2$\cite{shi2015superconductivity}. All studies have focused on the electron doped pairing, due to experimental limitations. For MoS$_2$, superconductivity is found to appear around $x_c=0.04-0.05$ electrons, reaching a maximum $T_c$ of $\sim 10.5$ K at $x\sim 0.1$\cite{ye2012superconducting}. In contrast, a few additional studies on MoS$_2$ and WS$_2$ have found $T_c$ to be relatively independent of doping, exhibiting a $x_c$ as low as $\sim 1\%$\cite{biscaras2015onset,jo2015electrostatically}. These results are in qualitative agreement with our calculated pairing instabilities, where we find a plateau in critical pairing strengths between $x=0.01$ and $0.02$ for MoS$_2$ (MoTe$_2$) and a sharp instability at $x=0.01$ for WS$2$. Reference \onlinecite{shi2015superconductivity} also suggests MoS$_2$ to produce the strongest superconducting state, followed by WS$_2$, MoSe$_2$ and MoTe$_2$. In contrast to experimental observations, we find a strong MoTe$_2$ instability, suggesting a possible competition between superconductivity and charge density wave state\cite{dong2018charge}. Recent scanning tunneling microscopy (STM) studies give some further insight to the symmetry of the gap functions. Specifically, measured local density of states displays a sharp V-like line shape, indicative of the nodal superconducting gap, inline with other suggestions of a mixed-parity state.  Moreover, high resolution real-space charge density maps show the presence of a pair-density wave, implying the Cooper pair has a finite center-of-mass momentum $Q$, thereby giving more credence to a unconventional origin of the pairing glue\cite{liu2021discovery,costanzo2018tunnelling}. Although we do not address Cooper pairing with a finite momentum in this work, our $Q=0$ results do indicate the prevalence of mix-parity states, with both nodal and fully gapped eigengap functions.\footnote{We note that for simplicity we have just considered the $Q=0$ case in this work, but our formalism can be straightforwardly extended to treat the pair density wave phase, which will be addressed in a future work. }  The non-monotonic decay in pairing strength for large electron doping illustrates the limits of a purely electronic (spin and charge fluctuation) driven pairing potential and suggests a cooperative feedback between electron and other bosonic modes, such as phonons, to generate this unconventional pairing state similar to that suggested in the cuprate high-temperature superconductors\cite{he2018rapid}.

Despite the lack of experiments for hole doped superconductivity in the TMDCs, we would like to point out some key features of our predictions in comparison to other theories. For dilute doping, our predicted gap functions are of mixed parity displaying both nodal and gap full solutions.  Here, the orbital character appears to be important, exhibiting a mixed interorbital and intraorbital pairing among $d_{x^2-y^2}$ and $d_{z^2}$ states, with higher order corrections between $d-p$ orbitals. These results are in contrast to those by Hsu {\it et al.},\cite{hsu2017topological} where they find chiral pairing states for both zero and finite center-of-mass momentum. Due to the single band nature of their model which intrinsically neglects the intertwining of strong $d-p$ hybridization and spin-orbit coupling, an accurate assessment of the gap symmetry is difficult to make\cite{kaba2019group}.  Moreover, the assumption of perfectly nesting circular Fermi surfaces, does not capture the anisotropy evident in the real materials, which will have knock-on effects on the anisotropy of the gap function along the Fermi surface.

Finally, we wish to discuss if topological superconductivity is expected in these TMDC compounds. To classify the various gap symmetries we have obtained as topologically trivial or non-trivial, we will concentrate on the odd-parity configurations. Generically, if no Fermi surface encloses a time reversible invariant momenta point $\mathbf{k}=\Gamma_i$ (where  $\Gamma_i=-\Gamma_i+\mathbf{G}$, and $\mathbf{G}$ is a reciprocal lattice vector) or each Fermi surface {\it reduces} to an $s$-wave pairing state, an odd-parity state leads to topological superconductivity\cite{sato2017topological}. For the electron doped states, since the Fermi pockets are centered on $K(K^\prime)$, and not $\Gamma$ or $M$, the nature of these states rests on their gap symmetry. States displaying $(s_i\pm s_o) s_z$ or $(s_i \pm s_o) f_z$ symmetries do not exhibit any nodes in the superconducting phase, rather a pocket dependent phase, implying they may be reduced to an $s$-wave pairing, implying these configurations to be topologically trivial. On the other hand, gap symmetries following $(d_i\pm d_o) s_z$, $(d_i\pm d_o)f_z$, $(s_i\pm i_o) s_z$,  and $(g_i\pm d_o) f_z$, and so on, clearly produce nodes in the superconducting state,  thus are classified as being topologically non-trivial. Moreover, many of these states are found to be non-degenerate, indicating the superconducting state should strongly break time-reversal symmetry. The same analysis holds for the lightly hole-doped results since they display the same Fermi-surface topology. Lastly, all $\Gamma$-centered hole pockets will produce non-trivial topological superconductivity. These states are predicted to be strongly time-reversal breaking, exhibiting large pairing strength eigenvalue value splitting of 0.01 for WS$_2$, with weaker, but finite, values of 0.003 in MoS$_2$.


\section{Concluding Remarks}\label{sec:conclusion}

We have developed a new methodology to tackle correlation-driven electron-electron pairing in material-specific detail. Using this effective potential on three prototypical  TMDC materials, we find a rich variety of pairing configurations exhibiting mixed parity. We find non-trivial topological superconductivity to be most prevalent and robust in hole-doped WS$_2$, with chiral modes spanning a wide range of hole carrier density. This initial study stands as a spring board for further detailed analysis of the pairing symmetry in TMDs and other 2D systems.

\begin{acknowledgments}
This work was carried out under the auspices of the US Department of Energy (DOE) National Nuclear Security Administration under Contract No. 89233218CNA000001. It was supported by the LANL LDRD Program, and in part by the Center for Integrated Nanotechnologies, a DOE BES user facility, in partnership with the LANL Institutional Computing Program for computational resources.
\end{acknowledgments}

\appendix
\section{Spin  and Orbital dependent Hedin's Equations}\label{Appendix:hedin}

Similar to Refs.~\onlinecite{hedin1965,aryasetiawan2008,lane2020interlayer}, the complete set of self-consistent orbital- and spin-dependent Hedin's equations relating the electronic self-energy $\Sigma$ to the Green's function $G$ and the screened interaction $W$, using the vertex $\Lambda$ and polarization function $P$ may be written as:
\begin{widetext}
\begin{subequations}
\begin{align}\label{eq:fullsetin}
\Sigma&_{\alpha n,\nu t}(1,5)=i\sigma_{\alpha\gamma}^{J}G_{\gamma k,\mu s}(1,4)  \Lambda^{L~ab}_{\mu s,\nu t}(4,5;6)W^{LJ}_{ak;bn}(6,1),\\
\nonumber\\
W&^{LJ}_{ak;bn}(6,1)= v^{ak;bn}_{LJ}(6,1) + v_{LM}^{ad;bc}(6,7)\chi_{0~cf;dg}^{MN}(7,8)W^{NJ}_{fk;gn}(8,1),\\
\nonumber\\
\chi&_{0~cf;dg}^{MN}(7,8)=-iG_{\delta c , \mu s}(7,9)\Lambda^{N~fg}_{\mu s,\nu t}(9,10;8) G_{\nu t , \xi d}(10,7^+)\sigma^{M}_{\xi \delta},\\
\nonumber\\
\Lambda&^{L~ab}_{\alpha n , \eta y}(1,4;6)=\delta(1,6)\delta(1,4)\sigma^{L}_{\alpha\eta }\delta_{an}\delta_{by}+\frac{\delta \Sigma_{\alpha n , \eta y}(1,4)}{\delta G_{\mu s , \nu t}(9,10)}G_{\nu t , \epsilon g}(9,11)\Lambda^{L~ab}_{\epsilon g , \delta f}(11,12;6)G_{\delta f , \mu s }(12,10).\label{eq:fullsetout}
\end{align}
\end{subequations}
\end{widetext}
To close the set of equations, Dyson's equation 
\begin{align}
G&_{\alpha n,\beta m}(1,2)=\\
&G_{0~\alpha n,\beta m}(1,2)+G_{0~\alpha n,\eta s}(1,3) \Sigma_{\eta s,\delta l}(3,4)G_{\delta l,\beta m}(4,2),\nonumber
\end{align}
links the fully interacting system to the bare non-interacting propagator,
\begin{align}
G^{-1}&_{0~\alpha n,\eta y}(1,4)=\\
&\left( i\frac{d}{dz_{1}} \delta_{\alpha\eta}\delta_{yn} - h^{0}_{\alpha n , \eta y}(1)-\Phi^{N}_{ny}(1)\sigma^{N}_{ \alpha \eta} \right)\delta(1,4),
\end{align}
where $\Phi^{J}$ is the total field, 
\begin{align}
\Phi^{J}_{nk}(1)=\pi^{J}_{nk}(1)+V^{J}_{H~k;n}(1),
\end{align}
$V_H$ is the Hartree potential 
\begin{align}
V^{J}_{H~k;n}(1)=\rho^{I}_{il}(3) v^{lk;in}_{IJ}(3,1),
\end{align}
and $\rho$ is the charge and spin density,
\begin{align}
\rho^{I}_{il}(3)=-iG_{\delta l,\xi i}(3,3^+) \sigma^{I}_{\xi \delta}.
\end{align}

Lastly, we note that the $GW$ approximation used in Eq.~(\ref{eq:sigma_gw}) is obtained by inserting bare vertex
\begin{align}
\Lambda&^{L~ab}_{\alpha n , \eta y}(1,4;6)= \delta(1,6)\delta(1,4)\sigma^{L}_{\alpha\eta }\delta_{an}\delta_{by}.
\end{align}
into Eq.~(\ref{eq:fullsetin}).

\section{Particle-Hole Propagator in the $\delta\rightarrow 0$ Limit for $\omega=0$}\label{Apendix:lindhard}
The bare polarization function contains a term of the form 
\begin{align}
\chi_{0}^{ij}(k,q,\omega)=\frac{  n_{F}^{j}(\mathbf{k}) - n_{F}^{i}(\mathbf{k+q})  }{ \omega + \Omega^{j}_{\mathbf{k}} - \Omega^{i}_{\mathbf{k+q}}  +i\delta}.
\end{align}
Taking the $\delta\rightarrow 0$ limit we can use the Sokhotski-Plemelj theorem for a Dirac delta function and obtain explicit forms for both the real and imaginary parts of $\chi_0$,
\begin{align}
\chi_{0}^{ij}(k,q,\omega)&=\mathcal{P}  \frac{  n_{F}^{j}(\mathbf{k}) - n_{F}^{i}(\mathbf{k+q})  }{ \omega + \Omega^{j}_{\mathbf{k}} - \Omega^{i}_{\mathbf{k+q}} } \\
&-
i\pi \left(  n_{F}^{j}(\mathbf{k}) - n_{F}^{i}(\mathbf{k+q})  \right) \delta(\omega + \Omega^{j}_{\mathbf{k}} - \Omega^{i}_{\mathbf{k+q}}).\nonumber
\end{align}
For $\omega=0$ the imaginary part is identically zero, leaving the real part as
\begin{align}
\chi_{0}^{ij}(k,q,\omega)&=\frac{  n_{F}^{j}(\mathbf{k}) - n_{F}^{i}(\mathbf{k+q})  }{  \Omega^{j}_{\mathbf{k}} - \Omega^{i}_{\mathbf{k+q}} } .
\end{align}
 Now, as the quasiparticle energies approach the Fermi level either as $\mathbf{q}\rightarrow 0$ or for specific nesting vectors $\mathbf{q}^*$, this expression becomes numerically unstable. To remedy this, we first re-write $\chi_0$ in terms of  
\begin{align}
\xi_{\mathbf{k,q}}^{ij}=\Omega^{j}_{\mathbf{k}}-\Omega^{i}_{\mathbf{k+q}}\\
\eta_{\mathbf{k,q}}^{ij}=\Omega^{j}_{\mathbf{k}}+\Omega^{i}_{\mathbf{k+q}}
\end{align}
yielding
\begin{align}
\frac{ -\sinh{\left(  \xi_{\mathbf{k,q}}^{ij} \beta/2  \right)}  }{  \cosh{\left(  \eta_{\mathbf{k,q}}^{ij} \beta/2  \right)}  + \cosh{\left(  \xi_{\mathbf{k,q}}^{ij} \beta/2  \right)}   }\frac{1}{ \xi_{\mathbf{k,q}}^{ij}} .
\end{align}
Clearly as $\xi_{\mathbf{k,q}}^{ij} \rightarrow 0$, $\sinh{ \left( \xi_{\mathbf{k,q}}^{ij}\beta/2 \right) } \approx  \xi_{\mathbf{k,q}}^{ij}\beta/2$, canceling the singularity. Therefore, in a small neighborhood about the Fermi level we explicitly expand $\chi_{0}^{ij}(k,q,\omega=0)$ as 
\begin{align}
&\frac{ -\frac{\beta}{2} \left(   1+  \frac{(\xi_{\mathbf{k,q}}^{ij} \beta/2)^1}{2!} + \frac{(\xi_{\mathbf{k,q}}^{ij} \beta/2)^2}{3!} +\dots \right)   }{  \cosh{\left(  \eta_{\mathbf{k,q}}^{ij} \beta/2  \right)}  + \cosh{\left(  \xi_{\mathbf{k,q}}^{ij} \beta/2  \right)}   }.
\end{align}
producing a stable numerical calculation. Now the only error introduced in the numerical evaluation of the bare susceptibility is the discretization of the momenta $\mathbf{k}$ and $\mathbf{q}$ on a finite grid.

\section{Numerically Stable Pairing Susceptibility}
The pairing susceptibility as written in Eq.~(\ref{eq:pairsus}) is not numerically stable as the band energies approach the Fermi level due to the effective singularity in the denominator. However, if we let 
\begin{subequations}
\begin{align}
\xi_{\mathbf{P}}^{xy}=\Omega^{x}_{\mathbf{P}}+\Omega^{y}_{-\mathbf{P}}\\
\eta_{\mathbf{P}}^{xy}=\Omega^{x}_{\mathbf{P}}-\Omega^{y}_{-\mathbf{P}}
\end{align}
\end{subequations}
we are able to recast Eq.~(\ref{eq:pairsus}) as
\begin{align}
\lambda^{P}_{xy} = \frac{\tanh{\left[ \left( \xi_{\mathbf{P}}^{xy}+\eta_{\mathbf{P}}^{xy} \right) \beta/4 \right]} + \tanh{\left[ \left( \xi_{\mathbf{P}}^{xy}-\eta_{\mathbf{P}}^{xy} \right) \beta/4 \right]}   }{\xi_{\mathbf{P}}^{xy}}.
\end{align}
Now, as the quasiparticle energies approach the Fermi level
\begin{align}
\tanh{\left[ \left( \xi_{\mathbf{P}}^{xy}+\eta_{\mathbf{P}}^{xy} \right) \beta/4 \right]} + \tanh{\left[ \left( \xi_{\mathbf{P}}^{xy}-\eta_{\mathbf{P}}^{xy} \right) \beta/4 \right]}\approx\\
\beta/4(1-\tanh{\left[ \eta_{\mathbf{P}}^{xy} \beta/4 \right]^{2}})\xi_{\mathbf{P}}^{xy}+\dots,\nonumber
\end{align}
 thus canceling the pole in the denominator. Therefore, in a small neighborhood about the Fermi level we approximate $\lambda^{P}_{xy}$ as 
\begin{align}
\lambda^{P}_{xy} &\approx \beta/4(1-\tanh{\left[ \eta_{\mathbf{P}}^{xy} \beta/4 \right]^{2}})
\end{align}
to allow for smooth numerical evaluation of the pairing susceptibility. This enables us to incorporate the temperature dependence of the superconducting gap without needing to formally introduce an arbitrary energy cut off.

\section{Superconducting Gap Symmetries for WS$_2$, MoTe$_2$, and MoS$_2$}
Tables~\ref{table:GapSymsWS2} - \ref{table:GapSymsMoTe2_2} give the gap symmetries for the first four pairing instabilities of 2H-WS$_2$, 2H-MoS$_2$, and 2H-MoTe$_2$, respectively for various hole and electron dopings at 10 K.

\begin{table}[h]
\footnotesize
\centering
\begin{tabular}{c|c|c|}
 $x$ & $\alpha$ & $\Delta^{\alpha}$  \\ \hline \hline
\multirow{3}{*}{-0.01}&$1-3$&$  \begin{array}{cl}   f (s_z\gamma^{0}+f_z\gamma^{z}) A^{d_{x^2-y^2}/d_{z^2}}_{asym}\\+ f (f_z\gamma^{0}+s_z\gamma^{z})  (A^{d_{x^2-y^2}}+A^{d_{z^2}})\end{array}$\\ \cline{2-3}
&$4$&$  \begin{array}{cl}  s (f_z\gamma^{0}+s_z\gamma^{z}) A^{d_{x^2-y^2}/d_{z^2}}_{asym}\\+ s (s_z\gamma^{0}+f_z\gamma^{z})  (A^{d_{x^2-y^2}}+A^{d_{z^2}})\end{array}$\\ \hline
\multirow{3}{*}{-0.02}&$1-2$&
$  \begin{array}{cl}  d (f_z\gamma^{0}+s_z\gamma^{z}) A^{d_{x^2-y^2}/d_{z^2}}_{asym}\\+ d (s_z\gamma^{0}+f_z\gamma^{z})  (A^{d_{x^2-y^2}}+A^{d_{z^2}})\end{array}$\\ \cline{2-3}
&$3-5$&
$ \begin{array}{cl}   f (s_z\gamma^{0}+f_z\gamma^{z}) A^{d_{x^2-y^2}/d_{z^2}}_{asym}\\+ f (f_z\gamma^{0}+s_z\gamma^{z})  (A^{d_{x^2-y^2}}+A^{d_{z^2}})\end{array}$\\ \hline
\multirow{3}{*}{-0.03}&$1-2$&
$  \begin{array}{cl}   \left[ d \gamma^{0} + (p \pm i p)(\gamma^{x}+\gamma^{y}) \right. \\
 \left. +p \gamma^{z}\right](A^{d_{xy}}+A^{d_{xy}/p_{y}}_{sym}) \end{array}$\\ \cline{2-3}
&$3-4$&
$ \begin{array}{ll}  \left[ (d+id) \gamma^{0} + (p \pm i p)(\gamma^{x}+\gamma^{y})  \right. \\ \left. +(p+ip) \gamma^{z}\right](A^{d_{xy}}+A^{d_{xy}/p_{y}}_{sym})\end{array}$\\  \hline
\multirow{2}{*}{-0.04}&$1-2$&
$ \begin{array}{cl}  \left[ d \gamma^{0} +p \gamma^{z}\right](A^{d_{xy}}+A^{d_{xy}/p_{y}}_{sym})\\+p \gamma^{x} A^{d_{xy}/p_{z}}_{sym}+p \gamma^{y} A^{d_{xy}/p_{x}}_{sym}\end{array}$\\ \cline{2-3}
&$3-6$& $ \begin{array}{cl}  (p+ip) \gamma^{z}   A^{d_{xy}/p_{z}}_{sym}
\\+(\gamma^{x}+\gamma^{y}) (p+ip) ( A^{d_{xy}/p_{y}}_{sym} + A^{d_{xy}} ) \end{array}$\\ \hline
\multirow{2}{*}{-0.05}&$1-2$&
$ \begin{array}{cl}    \left[(g_1+d_2) s_z \gamma^{0} + p s_z \gamma^{z}\right] (A^{d_{xy}/p_{y}}_{sym}  + A^{dxy}) \\+  p s_z \gamma^{x}A^{d_{xy}/p_{z}}_{sym} +  p s_z \gamma^{y}A^{d_{xy}/p_{x}}_{sym} \end{array}$\\ \cline{2-3}
&$3-4$&
$  \begin{array}{cl}  (p+ip) \gamma ^{z}  A^{d_{xy}/p_{z}}_{sym}
\\+(\gamma^{x}+\gamma^{y}) (p+ip) ( A^{d_{xy}/p_{y}}_{sym} + A^{d_{xy}} ) \end{array}$\\  
  \hline \hline
\multirow{4}{*}{0.01}&$1$&
$ \begin{array}{cl}  \left( (s_1-s_2)s_z \gamma^{0} + (s_1+s_2)f_z \gamma^{z} \right)   \left( A^{d_{xy}}+ A^{d_{xy}/p_{x}}_{sym} \right) \\  +  \left( (s_1-s_2)f_z \gamma^{0} + (s_1+s_2)s_z \gamma^{z} \right)  A^{d_{xy}/p_{z}}_{asym} 
\end{array} $\\ \cline{2-3}
&$2-3$&
$   (i_1+i_2) s_z \gamma^{0} A^{d_{xy}} + (i_1+i_2)f_z \gamma^{z} A^{d_{xy}}   $\\ \hline
\multirow{4}{*}{0.02}&$1$&
$ (s_1-s_2) s_z \gamma^{0} A^{d_{xy}} + (s_1+s_2)f_z \gamma^{z} A^{d_{xy}}    $\\ \cline{2-3}
&$2$&
$ (i_1-i_2) s_z \gamma^{0}A^{d_{xy}} + (i_1+i_2)f_z \gamma^{z}A^{d_{xy}}  $\\ \cline{2-3}
&$3-4$&
$ (d_1+d_2) s_z \gamma^{0} A^{d_{xy}} + (d_1-d_2)f_z \gamma^{z} A^{d_{xy}}  $\\  \hline
\multirow{5}{*}{0.03}&$1$&
$ (s_1-s_2) s_z \gamma^{0}A^{d_{xy}} + (s_1+s_2)f_z \gamma^{z}A^{d_{xy}} $\\ \cline{2-3}
&$2-3$&
$  (d_1+d_2) s_z \gamma^{0} A^{d_{xy}}+ (d_1-d_2)f_z \gamma^{z}  A^{d_{xy}}$\\ \cline{2-3}
&$4$&
$  (s_1-i_2) s_z \gamma^{0}  A^{d_{xy}}+ (s_1+i_2)f_z \gamma^{z} A^{d_{xy}}  $\\  \hline
\multirow{5}{*}{0.04}&$1$&
$ (s_1-s_2) s_z \gamma^{0}A^{d_{xy}} + (s_1+s_2)f_z \gamma^{z}A^{d_{xy}} $\\ \cline{2-3}
&$2-3$&
$  (d_1+d_2) s_z \gamma^{0}A^{d_{xy}} + (d_1-d_2)f_z \gamma^{z}A^{d_{xy}} $\\ \cline{2-3}
&$4$&
$ (g_1+d_2) s_z \gamma^{0}A^{d_{xy}} + (g_1-d_2)f_z \gamma^{z}A^{d_{xy}} $\\ \hline
\multirow{5}{*}{0.05}&$1$&
$  (s_1-s_2) s_z \gamma^{0} A^{d_{xy}}+ (s_1+s_2)f_z \gamma^{z}A^{d_{xy}} $\\ \cline{2-3}
&$2-3$&
$  (d_1 s_z+p_2 f_z) \gamma^{0}A^{d_{xy}} + (d_1 f_z+p_2 s_z) \gamma^{z}A^{d_{xy}} $\\ \cline{2-3}
&$4-5$&
$ (d_1+d_2) s_z \gamma^{0}A^{d_{xy}} + (d_1-d_2)f_z \gamma^{z}A^{d_{xy}} $\\  \hline
\end{tabular}
\caption{Predicted gap function symmetries for the leading four pairing instabilities in 2H-WS$_2$ under electron and hole doping at 10 K.}\label{table:GapSymsWS2}
\end{table}

\begin{table}[h]
\footnotesize
\centering
\begin{tabular}{c|c|c|}
 $x$ & $\alpha$ & $\Delta^{\alpha}$  \\ \hline \hline
\multirow{2}{*}{-0.01}&$1-2$&$ p\gamma^{0}(A^{d_{xy}} + A^{d_{xy}/p_y}_{sym} )$\\ \cline{2-3}
&$3-4$&$  (p+ip) (\gamma^{x}+\gamma^{y}) (  A^{d_{xy}}  + A_{sym}^{d_{xy}/p_y}) $\\ \hline
\multirow{2}{*}{-0.02}&$1-2$&$ p\gamma^{0}(A^{d_{xy}} + A^{d_{xy}/p_y}_{sym} )$\\ \cline{2-3}
&$3-4$&$  (p+ip) (\gamma^{x}+\gamma^{y}) (  A^{d_{xy}}  + A_{sym}^{d_{xy}/p_y}) $\\ \hline
\multirow{2}{*}{-0.03}&$1-2$&$ p\gamma^{0}(A^{d_{xy}} + A^{d_{xy}/p_y}_{sym} )$\\ \cline{2-3}
&$3-4$&$  (p+ip) (\gamma^{x}+\gamma^{y}) (  A^{d_{xy}}  + A_{sym}^{d_{xy}/p_y}) $\\ \hline
\multirow{2}{*}{-0.04}&$1-2$&$ p\gamma^{0}(A^{d_{xy}} + A^{d_{xy}/p_y}_{sym} )$\\ \cline{2-3}
&$3-4$&$  (p+ip) (\gamma^{x}+\gamma^{y}) (  A^{d_{xy}}  + A_{sym}^{d_{xy}/p_y}) $\\ \hline
\multirow{2}{*}{-0.05}&$1-2$&$ p\gamma^{0}(A^{d_{xy}} + A^{d_{xy}/p_y}_{sym} )$\\ \cline{2-3}
&$3-4$&$  (p+ip) (\gamma^{x}+\gamma^{y}) (  A^{d_{xy}}  + A_{sym}^{d_{xy}/p_y}) $\\ \hline \hline
\multirow{4}{*}{0.01}&$1$&
$ \begin{array}{cl}  
(s_1-s_2)s_z \gamma^{0} \left[ s_z \left(  A_{sym}^{d_{xy}/p_x}+A^{d_{xy}}  \right) + f_z\gamma^{0} A_{asym}^{d_{xy}/p_x} \right] \\ 
+(s_1+s_2)\gamma^{z}\left[ f_z  \left(  A_{sym}^{d_{xy}/p_x}+A^{d_{xy}}  \right) + s_z\gamma^{z} A_{asym}^{d_{xy}/p_z} \right] 
\end{array} $\\ \cline{2-3}
&$2-3$&
$ \begin{array}{cl}  
(s_1+s_2) \gamma^{x} \left[ f_z \left(  A_{sym}^{d_{xy}/p_x}+A^{d_{xy}}  \right) + s_z A_{asym}^{d_{xy}/p_x} \right] \\ 
+(s_1+s_2) \gamma^{y} \left[ f_z \left(  A_{sym}^{d_{xy}/p_x}+A^{d_{xy}}  \right) + s_z A_{asym}^{d_{xy}/p_z} \right] 
\end{array} $\\ \cline{2-3}
&$4$&
$ (d_1+d_2) \gamma^{0} \left[ s_z \left(  A_{sym}^{d_{xy}/p_x}+A^{d_{xy}}  \right) + f_z A_{asym}^{d_{xy}/p_z} \right] $\\ \hline
\multirow{4}{*}{0.02}&$1$&
$ \begin{array}{cl}   (s_1+s_2) s_z \gamma^{0} A^{d_{xy}} \\
+ (s_1+s_2) \gamma^{z} \left[ f_z (A^{d_{xy}}+A^{d_{xy}/p_x}_{sym} )+s_z A_{asym}^{d_{xy}/p_z} \right] \end{array} $
\\ \cline{2-3}
&$2-3$&
$ (s_1+s_2) (\gamma^{x}+\gamma^{y}) \left[ f_z (A^{d_{xy}}+A^{d_{xy}/p_x}_{sym} )+s_z A_{asym}^{d_{xy}/p_z} \right]  $\\ \cline{2-3}
&$4$&
$ \begin{array}{cl}   (f_1+f_2) f_z \gamma^{0} A^{d_{xy}} + (f_1 f_z+s_2 s_z) \gamma^{z} A_{asym}^{d_{xy}/p_z}\\ 
+ (f_1 s_z+s_2 f_z) \gamma^{z}  (A^{d_{xy}}+A^{d_{xy}/p_x}_{sym} )  \end{array} $\\  \hline
\multirow{5}{*}{0.03}&$1$&
$ \begin{array}{cl}   (s_1+s_2) s_z \gamma^{0} A^{d_{xy}} \\
+ (s_1+s_2) \gamma^{z} \left[ f_z (A^{d_{xy}}+A^{d_{xy}/p_x}_{sym} )+s_z A_{asym}^{d_{xy}/p_z} \right] \end{array} $
\\ \cline{2-3}
&$2-3$&
$ (s_1+s_2) (\gamma^{x}+\gamma^{y}) \left[ f_z (A^{d_{xy}}+A^{d_{xy}/p_x}_{sym} )+s_z A_{asym}^{d_{xy}/p_z} \right]  $\\ \cline{2-3}
&$4$&
$ \begin{array}{cl}   (f_1+f_2) f_z \gamma^{0} A^{d_{xy}} + (f_1 +f_2 ) f_z \gamma^{z} A_{asym}^{d_{xy}/p_z}\\ 
+ (f_1 + f_2) s_z \gamma^{z}  (A^{d_{xy}}+A^{d_{xy}/p_x}_{sym} )  \end{array} $\\  \hline
\multirow{5}{*}{0.04}&$1$&
$ \begin{array}{cl}   (s_1-s_2) s_z \gamma^{0} A^{d_{xy}} \\
+ (s_1+s_2) \gamma^{z} \left[ f_z (A^{d_{xy}}+A^{d_{xy}/p_x}_{sym} )+s_z A_{asym}^{d_{xy}/p_z} \right] \end{array} $
\\ \cline{2-3}
&$2-3$&
$ (s_1+s_2) (\gamma^{x}+\gamma^{y}) \left[ f_z (A^{d_{xy}}+A^{d_{xy}/p_x}_{sym} )+s_z A_{asym}^{d_{xy}/p_z} \right]  $\\ \cline{2-3}
&$4$&
$ \begin{array}{cl}   (d_1+d_2) s_z \gamma^{0} A^{d_{xy}} + (d_1 +d_2 ) s_z \gamma^{z} A_{asym}^{d_{xy}/p_z}\\ 
+ (d_1 + d_2) f_z \gamma^{z}  (A^{d_{xy}}+A^{d_{xy}/p_x}_{sym} )  \end{array} $\\  \hline
\multirow{5}{*}{0.05}&$1$&
$ \begin{array}{cl}   (s_1-s_2) s_z \gamma^{0} A^{d_{xy}} \\
+ (s_1+s_2) \gamma^{z} \left[ f_z (A^{d_{xy}}+A^{d_{xy}/p_x}_{sym} )+s_z A_{asym}^{d_{xy}/p_z} \right] \end{array} $
\\ \cline{2-3}
&$2-3$&
$ (s_1+s_2) (\gamma^{x}+\gamma^{y}) \left[ f_z (A^{d_{xy}}+A^{d_{xy}/p_x}_{sym} )+s_z A_{asym}^{d_{xy}/p_z} \right]  $\\ \cline{2-3}
&$4$&
$ \begin{array}{cl}   (d_1+d_2) s_z \gamma^{0} A^{d_{xy}} + (d_1 +d_2 ) s_z \gamma^{z} A_{asym}^{d_{xy}/p_z}\\ 
+ (d_1 + d_2) f_z \gamma^{z}  (A^{d_{xy}}+A^{d_{xy}/p_x}_{sym} )  \end{array} $\\  \hline
\end{tabular}
\caption{The same as Table~\ref{table:GapSymsWS2}, except for 2H-MoS$_2$.}\label{table:GapSymsMoS2}
\end{table}

\begin{table}[h]
\footnotesize
\centering
\begin{tabular}{c|c|c|}
 $x$ & $\alpha$ & $\Delta^{\alpha}$  \\ \hline \hline
\multirow{3}{*}{-0.01}&$1$&$  \begin{array}{cl}   d (f_z\gamma^{0}+s_z\gamma^{z}) A^{d_{x^2-y^2}/d_{z^2}}_{asym}\\+ d (s_z\gamma^{0}+f_z\gamma^{z})  (A^{d_{x^2-y^2}}+A^{d_{z^2}})\end{array}$\\ \cline{2-3}
&$2$&$  \begin{array}{cl}   f (s_z\gamma^{0}+f_z\gamma^{z}) A^{d_{x^2-y^2}/d_{z^2}}_{asym}\\+ f (f_z\gamma^{0}+s_z\gamma^{z})  (A^{d_{x^2-y^2}}+A^{d_{z^2}})\end{array}$\\ \cline{2-3}
&$3-4$&$  \begin{array}{cl}   g (f_z\gamma^{0}+s_z\gamma^{z}) A^{d_{x^2-y^2}/d_{z^2}}_{asym}\\+ g (s_z\gamma^{0}+f_z\gamma^{z})  (A^{d_{x^2-y^2}}+A^{d_{z^2}})\end{array}$\\ \hline
\multirow{3}{*}{-0.02}&$1$&$  \begin{array}{cl}   f (s_z\gamma^{0}+f_z\gamma^{z}) A^{d_{x^2-y^2}/d_{z^2}}_{asym}\\+ f (f_z\gamma^{0}+s_z\gamma^{z})  (A^{d_{x^2-y^2}}+A^{d_{z^2}})\end{array}$\\ \cline{2-3}
&$2$&$  \begin{array}{cl}   f (s_z\gamma^{0}+f_z\gamma^{z}) A^{d_{x^2-y^2}/d_{z^2}}_{asym}\\+ f (f_z\gamma^{0}+s_z\gamma^{z})  (A^{d_{x^2-y^2}}+A^{d_{z^2}})\end{array}$\\ \cline{2-3}
&$3-4$&$  \begin{array}{cl}   g (f_z\gamma^{0}+s_z\gamma^{z}) A^{d_{x^2-y^2}/d_{z^2}}_{asym}\\+ g (s_z\gamma^{0}+f_z\gamma^{z})  (A^{d_{x^2-y^2}}+A^{d_{z^2}})\end{array}$\\ \hline
\multirow{3}{*}{-0.03}&$1$&$  \begin{array}{cl}   f (s_z\gamma^{0}+f_z\gamma^{z}) A^{d_{x^2-y^2}/d_{z^2}}_{asym}\\+ f (f_z\gamma^{0}+s_z\gamma^{z})  (A^{d_{x^2-y^2}}+A^{d_{z^2}})\end{array}$\\ \cline{2-3}
&$2$&$  \begin{array}{cl}   f (s_z\gamma^{0}+f_z\gamma^{z}) A^{d_{x^2-y^2}/d_{z^2}}_{asym}\\+ f (f_z\gamma^{0}+s_z\gamma^{z})  (A^{d_{x^2-y^2}}+A^{d_{z^2}})\end{array}$\\ \cline{2-3}
&$3-4$&$  \begin{array}{cl}   g (f_z\gamma^{0}+s_z\gamma^{z}) A^{d_{x^2-y^2}/d_{z^2}}_{asym}\\+ g (s_z\gamma^{0}+f_z\gamma^{z})  (A^{d_{x^2-y^2}}+A^{d_{z^2}})\end{array}$\\ \hline
\multirow{3}{*}{-0.04}&$1$&$  \begin{array}{cl}   f (s_z\gamma^{0}+f_z\gamma^{z}) A^{d_{x^2-y^2}/d_{z^2}}_{asym}\\+ f (f_z\gamma^{0}+s_z\gamma^{z})  (A^{d_{x^2-y^2}}+A^{d_{z^2}})\end{array}$\\ \cline{2-3}
&$2$&$  \begin{array}{cl}   f (s_z\gamma^{0}+f_z\gamma^{z}) A^{d_{x^2-y^2}/d_{z^2}}_{asym}\\+ f (f_z\gamma^{0}+s_z\gamma^{z})  (A^{d_{x^2-y^2}}+A^{d_{z^2}})\end{array}$\\ \cline{2-3}
&$3-4$&$  \begin{array}{cl}   g (f_z\gamma^{0}+s_z\gamma^{z}) A^{d_{x^2-y^2}/d_{z^2}}_{asym}\\+ g (s_z\gamma^{0}+f_z\gamma^{z})  (A^{d_{x^2-y^2}}+A^{d_{z^2}})\end{array}$\\ \hline
\multirow{3}{*}{-0.05}&$1$&$  \begin{array}{cl}   f (s_z\gamma^{0}+f_z\gamma^{z}) A^{d_{x^2-y^2}/d_{z^2}}_{asym}\\+ f (f_z\gamma^{0}+s_z\gamma^{z})  (A^{d_{x^2-y^2}}+A^{d_{z^2}})\end{array}$\\ \cline{2-3}
&$2$&$  \begin{array}{cl}   f (s_z\gamma^{0}+f_z\gamma^{z}) A^{d_{x^2-y^2}/d_{z^2}}_{asym}\\+ f (f_z\gamma^{0}+s_z\gamma^{z})  (A^{d_{x^2-y^2}}+A^{d_{z^2}})\end{array}$\\ \cline{2-3}
&$3-4$&$  \begin{array}{cl}   g (f_z\gamma^{0}+s_z\gamma^{z}) A^{d_{x^2-y^2}/d_{z^2}}_{asym}\\+ g (s_z\gamma^{0}+f_z\gamma^{z})  (A^{d_{x^2-y^2}}+A^{d_{z^2}})\end{array}$\\ \hline
 \end{tabular}
\caption{The same as Table~\ref{table:GapSymsWS2}, except for hole doped 2H-MoTe$_2$.}\label{table:GapSymsMoTe2_1}
\end{table}
  
 \begin{table}[h]
\footnotesize
\centering
\begin{tabular}{c|c|c|}
 $x$ & $\alpha$ & $\Delta^{\alpha}$  \\ \hline \hline
\multirow{4}{*}{0.01}&$1$&
$ \begin{array}{cl}  
(s_1-s_2) \gamma^{0} \left[ s_z \left(  A_{sym}^{d_{xy}/p_x}+A^{d_{xy}}  \right) + f_z\gamma^{0} A_{asym}^{d_{xy}/p_x} \right] \\ 
+(s_1+s_2)\gamma^{z}\left[ f_z  \left(  A_{sym}^{d_{xy}/p_x}+A^{d_{xy}}  \right) + s_z\gamma^{z} A_{asym}^{d_{xy}/p_z} \right] 
\end{array} $\\ \cline{2-3}
&$2-3$&
$ \begin{array}{cl}  
(d_1+g_2) \gamma^{0} \left[ s_z \left(  A_{sym}^{d_{xy}/p_x}+A^{d_{xy}}  \right) + f_z\gamma^{0} A_{asym}^{d_{xy}/p_x} \right] \\ 
+(d_1+g_2)\gamma^{z}\left[ f_z  \left(  A_{sym}^{d_{xy}/p_x}+A^{d_{xy}}  \right) + s_z\gamma^{z} A_{asym}^{d_{xy}/p_z} \right] 
\end{array} $\\ \cline{2-3}
&$4$&
$ \begin{array}{cl}  
(s_1+i_2)\gamma^{0} \left[ s_z \left(  A_{sym}^{d_{xy}/p_x}+A^{d_{xy}}  \right) + f_z\gamma^{0} A_{asym}^{d_{xy}/p_x} \right] \\ 
+(s_1+i_2)\gamma^{z}\left[ f_z  \left(  A_{sym}^{d_{xy}/p_x}+A^{d_{xy}}  \right) + s_z\gamma^{z} A_{asym}^{d_{xy}/p_z} \right] 
\end{array} $\\ \hline
\multirow{4}{*}{0.02}&$1$&
$ \begin{array}{cl}  
(s_1-s_2)\gamma^{0} \left[ s_z \left(  A_{sym}^{d_{xy}/p_x}+A^{d_{xy}}  \right) + f_z\gamma^{0} A_{asym}^{d_{xy}/p_x} \right] \\ 
+(s_1+s_2)\gamma^{z}\left[ f_z  \left(  A_{sym}^{d_{xy}/p_x}+A^{d_{xy}}  \right) + s_z\gamma^{z} A_{asym}^{d_{xy}/p_z} \right] 
\end{array} $\\ \cline{2-3}
&$2-3$&
$ \begin{array}{cl}  
(d_1+d_2) \gamma^{0} \left[ s_z \left(  A_{sym}^{d_{xy}/p_x}+A^{d_{xy}}  \right) + f_z\gamma^{0} A_{asym}^{d_{xy}/p_x} \right] \\ 
+(d_1-d_2)\gamma^{z}\left[ f_z  \left(  A_{sym}^{d_{xy}/p_x}+A^{d_{xy}}  \right) + s_z\gamma^{z} A_{asym}^{d_{xy}/p_z} \right] 
\end{array} $\\ \cline{2-3}
&$4-5$&
$ \begin{array}{cl}  
(g_1-g_2) \gamma^{0} \left[ s_z \left(  A_{sym}^{d_{xy}/p_x}+A^{d_{xy}}  \right) + f_z\gamma^{0} A_{asym}^{d_{xy}/p_x} \right] \\ 
+(g_1+g_2)\gamma^{z}\left[ f_z  \left(  A_{sym}^{d_{xy}/p_x}+A^{d_{xy}}  \right) + s_z\gamma^{z} A_{asym}^{d_{xy}/p_z} \right] 
\end{array} $\\ \hline
\multirow{5}{*}{0.03}&$1$&
$ \begin{array}{cl}  
(s_1-s_2)\gamma^{0} \left[ s_z \left(  A_{sym}^{d_{xy}/p_x}+A^{d_{xy}}  \right) + f_z\gamma^{0} A_{asym}^{d_{xy}/p_x} \right] \\ 
+(s_1+s_2)\gamma^{z}\left[ f_z  \left(  A_{sym}^{d_{xy}/p_x}+A^{d_{xy}}  \right) + s_z\gamma^{z} A_{asym}^{d_{xy}/p_z} \right] 
\end{array} $\\ \cline{2-3}
&$2-3$&
$ \begin{array}{cl}  
(d_1+d_2) \gamma^{0} \left[ s_z \left(  A_{sym}^{d_{xy}/p_x}+A^{d_{xy}}  \right) + f_z\gamma^{0} A_{asym}^{d_{xy}/p_x} \right] \\ 
+(d_1-d_2)\gamma^{z}\left[ f_z  \left(  A_{sym}^{d_{xy}/p_x}+A^{d_{xy}}  \right) + s_z\gamma^{z} A_{asym}^{d_{xy}/p_z} \right] 
\end{array} $\\ \cline{2-3}
&$4$&
$ \begin{array}{cl}  
(f_1-f_2) \gamma^{0} \left[ f_z \left(  A_{sym}^{d_{xy}/p_x}+A^{d_{xy}}  \right) + s_z\gamma^{0} A_{asym}^{d_{xy}/p_x} \right] \\ 
+(f_1+f_2)\gamma^{z}\left[ s_z  \left(  A_{sym}^{d_{xy}/p_x}+A^{d_{xy}}  \right) + f_z\gamma^{z} A_{asym}^{d_{xy}/p_z} \right] 
\end{array} $\\ \hline
\multirow{5}{*}{0.04}&$1$&
$ \begin{array}{cl}  
(s_1-s_2)\gamma^{0} \left[ s_z \left(  A_{sym}^{d_{xy}/p_x}+A^{d_{xy}}  \right) + f_z\gamma^{0} A_{asym}^{d_{xy}/d_{z^2}} \right] \\ 
+(s_1+s_2)\gamma^{z}\left[ f_z  \left(  A_{sym}^{d_{xy}/p_x}+A^{d_{xy}}  \right) + s_z\gamma^{z} A_{asym}^{d_{xy}/d_{z^2}} \right] 
\end{array} $\\ \cline{2-3}
&$2-3$&
$ \begin{array}{cl}  
(d_1+d_2) \gamma^{0} \left[ s_z \left(  A_{sym}^{d_{xy}/p_x}+A^{d_{xy}}  \right) + f_z\gamma^{0} A_{asym}^{d_{xy}/d_{z^2}} \right] \\ 
+(d_1-d_2)\gamma^{z}\left[ f_z  \left(  A_{sym}^{d_{xy}/p_x}+A^{d_{xy}}  \right) + s_z\gamma^{z} A_{asym}^{d_{xy}/d_{z^2}} \right] 
\end{array} $\\ \cline{2-3}
&$4$&
$ \begin{array}{cl}  
(f_1-f_2) \gamma^{0} \left[ f_z \left(  A_{sym}^{d_{xy}/p_x}+A^{d_{xy}}  \right) + s_z\gamma^{0} A_{asym}^{d_{xy}/d_{z^2}} \right] \\ 
+(f_1+f_2)\gamma^{z}\left[ s_z  \left(  A_{sym}^{d_{xy}/p_x}+A^{d_{xy}}  \right) + f_z\gamma^{z} A_{asym}^{d_{xy}/d_{z^2}} \right] 
\end{array} $\\ \hline
\multirow{5}{*}{0.05}&$1$&
$ \begin{array}{cl}  
(s_1-s_2)\gamma^{0} \left[ s_z \left(  A_{sym}^{d_{xy}/p_x}+A^{d_{xy}}  \right) + f_z\gamma^{0} A_{asym}^{d_{xy}/d_{z^2}} \right] \\ 
+(s_1+s_2)\gamma^{z}\left[ f_z  \left(  A_{sym}^{d_{xy}/p_x}+A^{d_{xy}}  \right) + s_z\gamma^{z} A_{asym}^{d_{xy}/d_{z^2}} \right] 
\end{array} $\\ \cline{2-3}
&$2-3$&
$ \begin{array}{cl}  
(d_1+d_2) \gamma^{0} \left[ s_z \left(  A_{sym}^{d_{xy}/p_x}+A^{d_{xy}}  \right) + f_z\gamma^{0} A_{asym}^{d_{xy}/d_{z^2}} \right] \\ 
+(d_1-d_2)\gamma^{z}\left[ f_z  \left(  A_{sym}^{d_{xy}/p_x}+A^{d_{xy}}  \right) + s_z\gamma^{z} A_{asym}^{d_{xy}/d_{z^2}} \right] 
\end{array} $\\ \cline{2-3}
&$4$&
$ \begin{array}{cl}  
(f_1-f_2) \gamma^{0} \left[ f_z \left(  A_{sym}^{d_{xy}/p_x}+A^{d_{xy}}  \right) + s_z\gamma^{0} A_{asym}^{d_{xy}/d_{z^2}} \right] \\ 
+(f_1+f_2)\gamma^{z}\left[ s_z  \left(  A_{sym}^{d_{xy}/p_x}+A^{d_{xy}}  \right) + f_z\gamma^{z} A_{asym}^{d_{xy}/d_{z^2}} \right] 
\end{array} $\\ \hline
\end{tabular}
\caption{The same as Table~\ref{table:GapSymsWS2}, except for electron doped 2H-MoTe$_2$..}\label{table:GapSymsMoTe2_2}
\end{table}

\clearpage

\bibliography{pairing_Refs}

\end{document}